\documentclass[%
 reprint,
 twocolumn,
 amsmath,amssymb,
 aps,
]{revtex4-2}

\usepackage{graphicx}% Include figure files
\usepackage{dcolumn}% Align table columns on decimal point
\usepackage{bm}% bold math

\usepackage{color}

\usepackage[hidelinks]{hyperref}
\usepackage{cleveref} 
\crefname{figure}{Figure}{Figures}
\Crefname{figure}{Figure}{Figures}

\begin{document}

\preprint{APS/123-QED}

\title{Quantum metrology with linear Lie algebra parameterisations}%

\author{Ruvi Lecamwasam}
 \email{me@ruvi.blog}
\affiliation{A*STAR Quantum Innovation Centre (Q.InC), Institute for Materials Research and Engineering (IMRE), Agency for Science, Technology and Research (A*STAR), 2 Fusionopolis Way, 08-03 Innovis 138634, Republic of Singapore}
\affiliation{Quantum Machines Unit, Okinawa Institute of Science and Technology Graduate University, Onna, Okinawa 904-0495, Japan}
\author{Tatiana Iakovleva}%
\affiliation{Quantum Machines Unit, Okinawa Institute of Science and Technology Graduate University, Onna, Okinawa 904-0495, Japan}
\author{Jason Twamley}%
\affiliation{Quantum Machines Unit, Okinawa Institute of Science and Technology Graduate University, Onna, Okinawa 904-0495, Japan}

\date{\today}

\begin{abstract}
Lie algebraic techniques are powerful and widely-used tools for studying dynamics and metrology in quantum optics. 
When the Hamiltonian generates a Lie algebra with finite dimension, the unitary evolution can be expressed as a finite product of exponentials using the Wei-Norman expansion. 
The system is then exactly described by a finite set of scalar differential equations, even if the Hilbert space is infinite. 
However, the differential equations provided by the Wei-Norman expansion are nonlinear and often have singularities that prevent both analytic and numerical evaluation. We derive a new Lie algebra expansion for the quantum Fisher information, which results in linear differential equations. 
Together with existing Lie algebra techniques this allows many metrology problems to be analysed entirely in the Heisenberg picture. This substantially reduces the calculations involved in many metrology problems, and provides analytical solutions for problems that cannot even be solved numerically using the Wei-Norman expansion. It also allows us to study general features of metrology problems, valid for all quantum states. We provide detailed examples of these methods applied to problems in quantum optics and nonlinear optomechanics. 

\end{abstract}

\maketitle

\section{Introduction}\label{sec:Introduction}
One of the most widely used tools for studying quantum sensing is the quantum Fisher information (QFI) \cite{meyer_fisher_2021,liu_quantum_2020,hayashi_quantum_2017,petz_introduction_2010}. Suppose a system undergoes evolution which depends on a parameter, such as gravity or a magnetic field. The QFI measures the smallest fluctuation in the parameter that can be detected from a measurement on the system, in a relation called the quantum Cramér-Rao bound \cite{helstrom_quantum_1969,holevo_probabilistic_2011,braunstein_statistical_1994}. 
%An important assumption of this theory is that the fluctuations occur around a well-known operating point, and in cases where this does not hold there are other methods of quantifying sensitivity \cite{hall_does_2012,demkowicz-dobrzanski_multi-parameter_2020}. However, the quantum Fisher information has played and continues to play a central role in metrology \cite{liu_quantum_2020,meyer_fisher_2021}. 
There are also diverse applications beyond sensing, examples include quantum resource theories \cite{tan_fisher_2021}, phase transitions \cite{gu_fidelity_2010,wang_quantum_2014,marzolino_fisher_2017}, entanglement \cite{toth_multipartite_2012,hyllus_fisher_2012}, thermodynamics \cite{marvian_operational_2022}, and even quantum error correction \cite{kubica_using_2021}.

These many connections likely stem from the relationship between quantum Fisher information and the fundamental geometry of quantum states \cite{braunstein_statistical_1994,sidhu_geometric_2020,liu_quantum_2020}. Suppose $\theta$ is the parameter we wish to measure, and $\rho_{\theta}$ is the quantum state resulting from the evolution. Our ability to detect a small fluctuation $\Delta\theta$ depends on our ability to differentiate $\rho_{\theta}$ and $\rho_{\theta+\Delta\theta}$. The quantum Fisher information is equal to the distance between these states, as measured by the Bures metric on the space of density matrices \cite{braunstein_statistical_1994}. We will consider unitary evolution $\rho_{\theta}=U_{\theta}\rho_0U^{\dagger}_{\theta}$ for some initial probe state $\rho_0$. If we picture $U_{\theta}$ as a curve parameterised by $\theta$ on the manifold of unitary matrices, the quantum Fisher information at $\theta$ then corresponds to the tangent vector to this curve at $U_{\theta}$. 
%Equivalently if we imagine $\theta$ as parameterising a curve $\rho_{\theta}$ on the manifold of possible states, we can find the QFI by computing the tangent vector at a point $\rho_{\theta}$ on this manifold, pointing in the direction the state would travel as $\theta$ increases.

Quantum dynamics are naturally described by the mathematical structure of a Lie algebra. 
This is a set of operators which satisfies two properties. The first is that the Lie algebra is a vector space, thus each element can be expressed as a linear combination of some basis of operators. The second property is that the commutator of any two elements of the algebra is also an element of the algebra. Given a Hamiltonian, we can generate a basis for its corresponding Lie algebra by beginning with the set of all operators in the Hamiltonian, expanding this by all new operators that arise from taking commutators of operators in the set, and then all possible commutators of the resulting operators, and so on \footnote{The Hamiltonian $H$ may be decomposed as a sum of operators in different ways. Each such decomposition provides a different basis for the Lie algebra. Different bases may be more or less efficient, which we explore in \cref{sec:ExampleSpin}.}. The resulting set of operators is called the `dynamical Lie algebra'. 

The basis for the dynamical Lie algebra may be infinite, meaning that we can continue to generate more operators by taking commutators, which are not expressible as a linear combination of existing basis elements. In many cases however, we eventually find a finite basis containing all the operators of the Hamiltonian, where the commutator of any two elements is a linear combination of operators in the basis. In this case the dynamics take place in a \emph{finite Lie algebra}, which can be leveraged to greatly simplify analysis.

Expressing operators in a basis adapted to the Lie algebra can make computations easier, as was used in \cite{warszawski_solving_2020} to derive a stochastic master equation. The most common application of finite Lie algebras however is the Wei-Norman expansion \cite{wei_global_1964,altafini_use_2002,qvarfort_solving_2022}, which uses the algebra's structure to parameterise the dynamics. If the Lie algebra basis contains $n$ elements $\{X_1,\ldots,X_n\}$, the Wei-Norman expansion expresses the unitary time-evolution operator as a product of $n$ exponentials $U_{\theta}(t)=e^{c_1(\theta,t)X_1}\cdots e^{c_n(\theta,t)X_n}$, with scalar coefficients $c_j$. Even if the quantum dynamics take place in an infinite-dimensional Hilbert space, we can derive $n$ coupled differential equations for the scalars $c_j$ which will exactly describe the evolution.

The Wei-Norman expansion is a standard tool in quantum optics \cite{ban_lie-algebra_1993,puri_mathematical_2001,qvarfort_solving_2022}. It has been extensively applied to study dynamics and calculate Fisher information in nonlinear optomechanics \cite{mancini_ponderomotive_1997,bose_preparation_1997,paredes-juarez_lie_2020,schneiter_optimal_2020,qvarfort_constraining_2022,qvarfort_time-evolution_2020}, and non-Markovian systems \cite{qvarfort_enhanced_2023}. We also note an approach based on the Magnus expansion, which can be seen as the `dual' to the Wei-Norman solution \cite{galitski_quantum--classical_2011}. The formalism of Liouville space \cite{ban_lie-algebra_1993,gyamfi_fundamentals_2020}, also known as `vectorisation', allows open system dynamics to be expressed in a manner analogous to unitary dynamics. Using this the Wei-Norman expansion can be applied to dissipative evolution \cite{scopa_exact_2019,teuber_solving_2020,qvarfort_master-equation_2021}. The Wei-Norman expansion is also used extensively in quantum computing and machine learning, as it provides a natural parameterisation for quantum circuits \cite{cerezo_variational_2021,wierichs_symmetric_2023,goh_lie-algebraic_2023,garcia-martin_architectures_2024}.

There are however some problems with the Wei-Norman expansion. The differential equations for the coefficients are nonlinear, making analysis difficult. Deriving these equations typically involves arduous algebra. The equations also depend on the ordering of the exponentials, but one generally does not know in advance which ordering will be easiest to solve. More troubling, there are often singularities in these equations which can prevent even numerical evaluation. These can sometimes be removed by analytical means \cite{altafini_use_2002, guerrero_semiclassical_2020}, but this is not possible in general. The singularities are analogous to those that arise in polar coordinates or Euler angles, the result of us forcing a particular coordinate system onto dynamics which have their own intrinsic geometry. It is thus natural to ask what other coordinate systems can be found which leverage the Lie algebra structure, but are better adapted to the quantum dynamics.

One alternative is provided by the work of Fernández, who showed how a finite Lie algebra could be used to parameterise Heisenberg-picture dynamics \cite{fernandez_time-evolution_1989}. The time-evolution of an element $X_j$ of the Lie algebra basis is expressed as $X_j(t)=c_{j1}(\theta,t)X_1+\cdots+c_{jn}(\theta,t)X_n$. Here the differential equations for the scalar coefficients $c_{jk}$ are linear. There is a price to be paid, in that there are now $n^2$ differential equations (though this number can often be reduced by exploiting the algebra's structure). However, linearity greatly simplifies analytical treatment. The equations are singularity-free, expanding the range of systems and parameter regimes that can be analysed and simulated. Heisenberg-picture equations for the operators can also be much more illuminating than the product of operator exponentials provided by the Wei-Norman expansion.

The approach of Fernández is much less known than the Wei-Norman method. It has been applied to study the dynamics of time-dependent quadratic systems with a phenomenological model of dissipation \cite{twamley_quantum_1993}, and more recently Bose-Einstein condensates \cite{haine_multi-mode_2005,haine_outcoupling_2005}. However as of yet, this method is not used by the metrology community. One probable reason is that currently, it isn't clear how to compute the quantum Fisher information in this formalism. It is still necessary to compute the time-evolution operator $U_{\theta}$ and differentiate this to obtain the tangent vector, or solve for the evolution of the quantum states $\rho_{\theta}$ and $\rho_{\theta+\Delta\theta}$ then compute the distance betweenthem. 

\begin{figure}
    \includegraphics[width=\columnwidth]{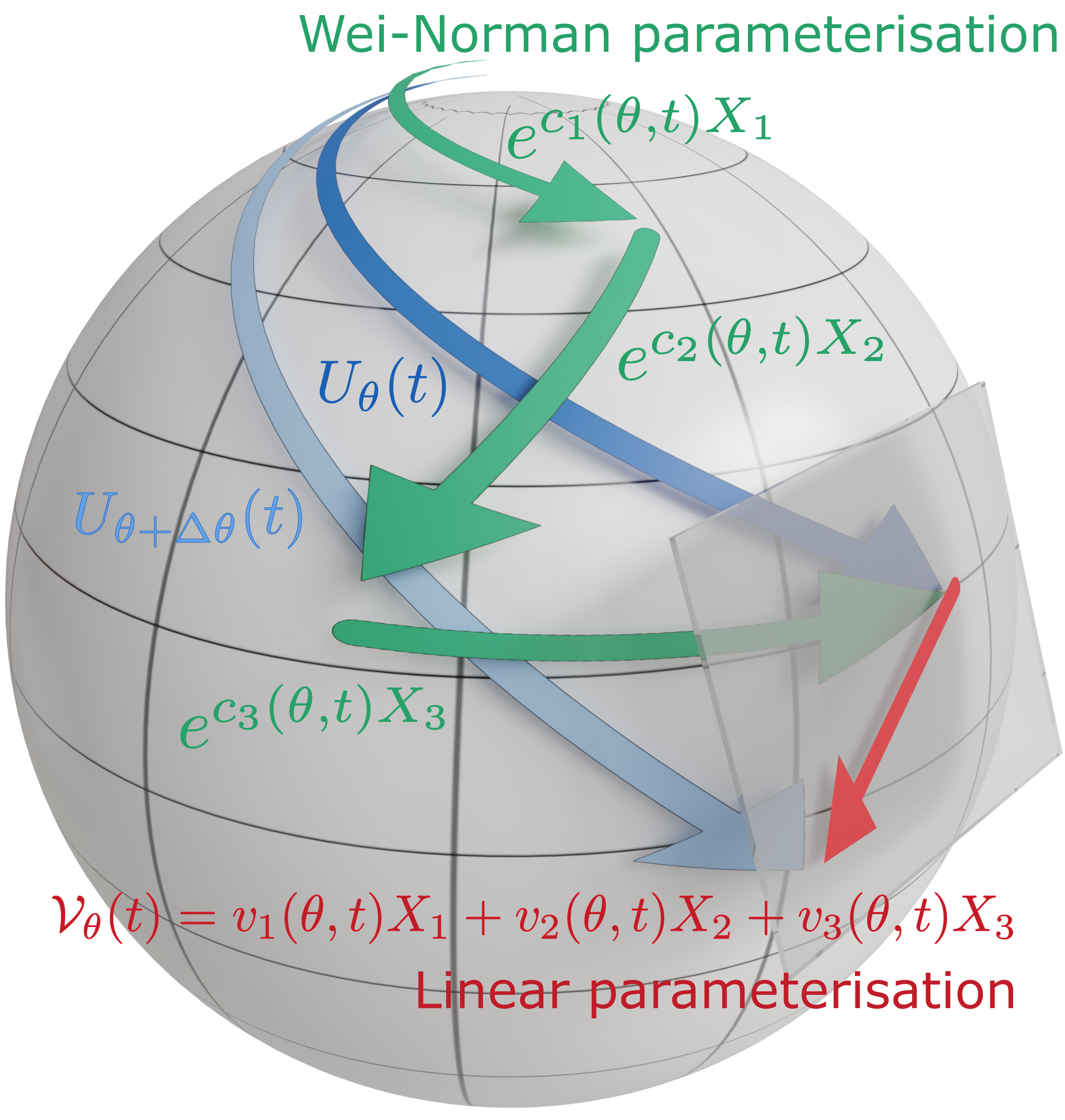}
    \caption{The standard Wei-Norman and our linear parameterisations for the quantum Fisher information (QFI). The system undergoes unitary evolution $U_{\theta}(t)$ which depends on a parameter $\theta$. The QFI quantifies our ability to measure infinitesimal fluctuations $\Delta\theta$ in the parameter, and corresponds to the tangent vector $\mathcal{V}_{\theta}$ pointing from $U_{\theta}(t)$ towards $U_{\theta+\Delta\theta}(t)$. The Wei-Norman expansion represents the evolution as a product of exponentials $U_{\theta}(t)=e^{c_1(\theta,t)X_1}\cdots e^{c_n(\theta,t)X_n}$, where the operators $X_j$ form a basis for Lie algebra corresponding to the dynamics. Differentiating this product with respect to $\theta$ then gives the QFI vector $\mathcal{V}_{\theta}$.
    %However the differential equations for the scalar coefficients $c_j$ are nonlinear, and have singularities which can prevent even numerical evaluation. Once these are solved, we must still compute the product of exponentials, and then its derivative. 
    Our method parameterises $\mathcal{V}_{\theta}$ directly, as a linear combination of the Lie algebra basis. This is possible because the space of tangent vectors is always linear, for example, the tangent plane to a sphere.
    %The $n$ equations for the scalar coefficients $v_j(\theta,t)$ are now linear. Thus these are often easier to solve, do not exhibit singularities, and we no longer have to compute and then differentiate the unitary operator $U_{\theta}$.
    }\label{fig:ParameterisationMethod}
\end{figure}

In this manuscript, we derive a linear Lie algebra parameterisation for the quantum Fisher information. Our method works by directly parameterising the tangent vector to $U_{\theta}$, as we illustrate in \cref{fig:ParameterisationMethod}.
When the dynamics take place in an $n$-dimensional Lie algebra, this results in $n$ linear scalar differential equations. 
%Our method applies to both time-independent, and time-dependent Hamiltonians. 
Together with the linear parameterisation of \cite{fernandez_time-evolution_1989}, this now allows metrological problems on finite Lie algebras to be analysed entirely in the Heisenberg picture, using linear parameterisations.  

In \cref{sec:FisherExpansions} we introduce our linear parameterisation for both time-independent and time-dependent Hamiltonians.
To illustrate this, we consider in \cref{sec:IllustrativeExamples} two simple examples: a two-level system, and an optical cavity. These show how linear parametersiations simplify calculation, and can succeed when the Wei-Norman expansion fails. We then in \cref{sec:ExampleOptomechanics} consider metrology with nonlinear optomechanics. Linear parameterisations allow us to derive simple, state-independent results that  Finally in \cref{sec:Discussion} 
we discuss possible avenues of future research.

\section{Linear parameterisation for the quantum Fisher information}\label{sec:FisherExpansions}
Suppose a state $\rho_0$ undergoes unitary evolution $U_{\theta}$ which depends on a parameter $\theta$, resulting in a quantum state $\rho_{\theta}=U_{\theta}\rho_0U_{\theta}^{\dagger}$. Quantum Fisher information asks the question, what is the smallest fluctuation in the parameter $\Delta\theta$ that we can detect, from a measurement on the final state. This will be determined by how different the states $\rho_{\theta}$ and $\rho_{\theta+\Delta\theta}$ are. This is quantified by the vector $\mathcal{V}_{\theta}$ pointing from $U_{\theta}$ to $U_{\theta+\Delta\theta}$ \cite{liu_quantum_2015} \footnote{Strictly speaking, the red arrow in the tangent plane depicted in \cref{fig:ParameterisationMethod} corresponds to the derivative $\partial_{\theta}U_{\theta}$. Multiplication by $U^\dagger_{\theta}$ in \cref{eq:QuantumFisherVector} then translates this vector to the upper pole of the sphere. This translation is necessary in order to compare vectors at different points. In mathematical terms, the Lie algebra of a Lie group is defined as the tangent space at the origin.}:
\begin{equation}\label{eq:QuantumFisherVector}
    \mathcal{V}_{\theta}= i\left(\partial_{\theta}U_{\theta} \right)U_{\theta}^{\dagger}.
\end{equation}
Note that in the case $U_{\theta} = e^{-iG\theta}$, where $G$ is a Hermitian operator encoding the parameter, we have $\mathcal{V}_{\theta}=G$. The expression \cref{eq:QuantumFisherVector} however is more general, and is valid for any form of $\theta$-dependence. 

The quantum Cramér-Rao theorem states that the variance in our estimate of $\theta$ is bounded by $\mathrm{var}(\theta)\ge1/\left(N\cdot \mathrm{QFI}(\theta)\right)$, where $N$ is the number of measurements, and $\mathrm{QFI}(\theta)$ is the quantum Fisher information. If the initial state $\rho_0$ is pure, the quantum Fisher information is given by the variance of $\mathcal{V}_{\theta}$ with respect to $\rho_0$: 
\begin{equation}\label{eq:QFIPure}
    \mathrm{QFI}(\theta)=4\,\mathrm{var}_{\rho_0}(\mathcal{V}_{\theta}).
\end{equation}
If $\rho_0$ is mixed, \cref{eq:QFIPure} gives an upper bound on the quantum Fisher information. There is an exact formula in the mixed state case, which involves decomposing $\rho_0$ into its eigenbasis \cite{liu_quantum_2015,schneiter_optimal_2020}. In \cref{app:Variance} we discuss how \cref{eq:QFIPure} may be efficiently computed.

Let us suppose our Hamiltonian $H$ generates a finite Lie algebra, with basis $\{X_1,\ldots,X_n\}$. The Hamiltonian can then be expanded as
\begin{equation}\label{eq:ParameterDependentHam}
    H=h_j(\theta)X_j,
\end{equation}
where $h_j$ are scalar coefficients which depend on $\theta$, and we use the convention of implicit summation over the repeated index $j$. Since the operators form a Lie algebra we can express any commutator as
\begin{equation}\label{eq:CommutatorGamma}
    [X_j,X_k]=\Gamma_{jk}^mX_m,
\end{equation}
where $\Gamma_{jk}^m$ are scalar coefficients, and we sum over the repeated index $m$. In the following sections we will exploit the Lie algebra structure to find simple formulae for the tangent vector $\mathcal{V}$. From now on we will omit subscripts $\theta$, and leave the parameter dependence implicit.
%Traditionally, this is done by first explicitly solving for $U$ using the Wei-Norman expansion, and then computing the derivative 
%in \cref{eq:QuantumFisherVector}. We will show however that $\mathcal{V}$ can be computed explicitly, in what is often a much more efficient process. 

\subsection{Time-independent Hamiltonian}\label{sec:FisherFormulaDerivation}
Let us first consider the case of a time-independent Hamiltonian $H$. The unitary evolution operator is
\begin{equation}
    U =e^{-iH t/\hbar}.
\end{equation}
Then from \cref{eq:QuantumFisherVector}, the quantum Fisher information vector is
\begin{equation}\label{eq:VUnitaryTimeIndependentH}
 \mathcal{V} = i\left(\partial_{\theta}e^{-iH t/\hbar}\right)e^{iH t/\hbar}.
\end{equation}
%We will evaluate this, using the fact that the Hamiltonian operators $X_j$ form a finite Lie algebra.

Let us begin with the derivative. For a scalar function $\alpha({\theta})$, we have  $ \partial_{\theta}e^{\alpha({\theta})} = \left[ \partial_{\theta}\alpha(\theta) \right] e^{\alpha(\theta)}$.
If $\alpha$ is an operator however, then it may not commute with its derivative, and attention needs to be payed to the ordering of $\partial_{\theta}\alpha(\theta)$ and $e^{\alpha(\theta)}$. 
%Let's derive a formula for this in the case of Lie algebra evolution. Consider the derivative $\partial_{\theta}e^{\theta\alpha+\beta}$. For scalars $\alpha,\beta$ this is $\alpha e^{\theta\alpha+\beta}$. When $\alpha$ and $\beta$ do not commute however, we have to take into account operator ordering. 
For an operator $A_{\theta}$ depending on $\theta$, the derivative is given by \cite[\S 2]{wilcox_exponential_1967}
\begin{equation}
    \partial_{\theta}e^{-A_{\theta}t}=-\int_0^te^{-(t-u)A_{\theta}}(\partial_{\theta}A_{\theta})e^{-uA_{\theta}}\mathrm{d}u.
    %\left[\frac{d}{d\theta}e^{H}\right]_{\theta=0}=\int_0^1\mathrm{d}s\,e^{s\alpha}\beta e^{\alpha (1-s)}.
\end{equation}
Setting $A_{\theta}=iH/\hbar$ and changing variables in the integrand to $s=t-u$ gives:
\begin{equation}
    \partial_{\theta}e^{-iH t/\hbar}=-\frac{i}{\hbar}\int_0^te^{-isH/\hbar}(\partial_{\theta}H )e^{i(s-t)H/\hbar }\,\mathrm{d}s.
\end{equation}
Then from \cref{eq:VUnitaryTimeIndependentH} the quantum Fisher information vector is
\begin{equation}\label{eq:VExpandedTimeIndependent}
    \mathcal{V} = \frac{1}{\hbar}\int_0^te^{-isH/\hbar}(\partial_{\theta}H )e^{isH/\hbar}\,\mathrm{d}s.
\end{equation}
Note that we have not yet used the fact that the Hamiltonian operators form a finite Lie algebra. Thus this formula is valid for a general time-independent Hamiltonian.

Let us now leverage the structure of the Lie algebra. We will make use of the Hadamard lemma:
\begin{equation}\label{eq:HadamardLemma}
    e^{\alpha X}Ye^{-\alpha X}=e^{\alpha\,\mathrm{ad}_X}Y.
\end{equation}
Here $\mathrm{ad}_X$ is the adjoint representation of the operator $X$. In an $n$-dimensional Lie algebra this is an $n\times n$ matrix, which can be computed from the commutation relations. We provide some background on this in the supplementary material. Substituting the Hadamard lemma into \cref{eq:VExpandedTimeIndependent} gives
\begin{equation}
    \mathcal{V} =\frac{1}{\hbar}\int_0^te^{-(i/\hbar)s\,\mathrm{ad}_{H}}(\partial_{\theta}H )\,\mathrm{d}s.
\end{equation}
%This expression is true whether or not the Lie algebra is finite. However, it is only in the case of a finite Lie algebra that the adjoint representation $\mathrm{ad}_{H }$ can be written as a finite-dimensional matrix. 
Substituting the Hamiltonian $H =h_j(\theta)X_j$ gives
\begin{equation}
    \mathcal{V} =\frac{1}{\hbar}\int_0^te^{-(i/\hbar)s\,h_k(\theta)\mathrm{ad}_{X_k}}(\partial_{\theta}h_j(\theta)X_j)\,\mathrm{d}s,
\end{equation}
which can be re-written as
\begin{equation}\label{eq:FisherGeneratorGeneralForm}
    \mathcal{V} = \partial_{\theta}h_j(\theta)\frac{1}{\hbar}\left[\int_0^t\exp\left(-\frac{i}{\hbar}s\,h_k(\theta)\mathrm{ad}_{X_k}\right)\,\mathrm{d}s\right]X_j.
\end{equation}
This gives an explicit formula for the quantum Fisher information vector where the Hamiltonian operators form a finite Lie algebra. The integrand is an $n\times n$ matrix, where $n$ is the dimension of the Lie algebra. Since $\mathrm{ad}_{X_k}X_j$ is a linear combination of Lie algebra elements, a consequence of \cref{eq:FisherGeneratorGeneralForm} is that the tangent vector $\mathcal{V}$ is also an element of the Lie algebra.

\subsection{Time-dependent Hamiltonian}\label{sec:FisherDerivationTimeDependent}
We will now consider a Hamiltonian which depends on time. Our strategy is to study the time evolution of the Fisher information vector over the course of the interaction. At a time $t$ the vector is given by:
\begin{equation}
    \mathcal{V}(t)=i\left[\partial_{\theta}U (t)\right]U ^{\dagger}(t).
\end{equation}
Let us differentiate this with respect to time. To evaluate this derivative, we will use commutation of partial derivatives: $\partial_t\partial_{\theta}X=\partial_{\theta}\partial_tX$ for any operator $X$. The time evolution operator $U(t)$ follows the Schr\"odinger equation $\partial_tU(t)=-(i/\hbar)H(t)U(t)$, and the conjugate expression $\partial_tU^{\dagger}(t)=(i/\hbar)U^{\dagger}(t)H(t)$. Putting these together gives:
\begin{equation}
    \begin{aligned}
        -i\hbar\partial_t\mathcal{V} &= \hbar\left(\partial_{\theta}\partial_tU \right)U ^{\dagger}+\hbar(\partial_{\theta}U )(\partial_tU ^{\dagger}), \\
            &= \left(\partial_{\theta}\left[-iH U \right]\right)U ^{\dagger}+\left(\partial_{\theta}U \right)\left(iU ^{\dagger}H \right), \\
            &= -i\left(\left(\partial_{\theta}H \right)U +H \partial_{\theta}U \right)U ^{\dagger}+i\left[(\partial_{\theta}U )U ^{\dagger}\right]H , \\
            &= -i\partial_{\theta}H -i\left[H ,(\partial_{\theta}U )U ^{\dagger}\right].
    \end{aligned}
\end{equation}
Thus the evolution of the Fisher information vector is:
\begin{equation}\label{eq:VHeisenbergEvolution}
    i\hbar\,\partial_t\mathcal{V}=[H,\mathcal{V}]+i\,\partial_{\theta}H.
    %i\hbar\,\partial_t\mathcal{V}=i\,\partial_{\theta}H -[\mathcal{V},H ].
\end{equation}
The initial condition is $\mathcal{V}(0)=0$. Since have not yet used any property of Lie algebras, \cref{eq:VHeisenbergEvolution} is valid for general unitary evolution. 

The evolution given by \cref{eq:VHeisenbergEvolution} will cause $\mathcal{V}$ to remain within the Lie algebra. This follows because since $H$ is an element of the Lie algebra, and Lie algebras are closed under commutators, both terms in the right-hand side are elements of the Lie algebra. 
Thus there exist scalars $v_j(t)$ such that $\mathcal{V}=v_j(t)X_j$. The Hamiltonian is $H =h_j(\theta,t)X_j$, for scalar $h_j(\theta,t)$. Substituting these into \cref{eq:VHeisenbergEvolution} then gives
\begin{equation}
    \dot{v}_j(t)X_j = \left(\partial_{\theta}h_j(\theta,t)X_j+iv_j(t)\left[X_j,H \right]\right)/\hbar,
\end{equation}
where we use a dot to denote the time-derivative. The commutator $[X_j,H]$ can be expanded as
\begin{equation}
    [X_j,H]=h_k(\theta,t)[X_j,X_k]=h_k(\theta,t)\Gamma_{jk}^lX_l,
\end{equation}
where $\Gamma_{jk}^l$ are the coefficients from \cref{eq:CommutatorGamma}. Equating coefficients of $X_j$ then gives
\begin{equation}\label{eq:TimeDependentFisherGenerator}
    i\hbar \,\dot{v}_j(t)=i\partial_{\theta}h_j(\theta,t)-\Gamma_{kl}^jh_l(\theta,t)v_k(t).
\end{equation}
This equation describes the time-evolution of the Fisher information generator. We can see that this is linear in the scalar coefficients $v_j(t)$. The initial condition $\mathcal{V}(0)=0$ then corresponds to:
\begin{equation}
    v_j(0)=0.
\end{equation}

%If the Hamiltonian is time-dependent, issues of operator ordering mean we can no longer find an explicit form for $\mathcal{V}$. Instead, we derive in \cref{sec:FisherDerivationTimeDependent} a differential equation describing its time evolution:
%\begin{equation}\label{eq:VHeisenbergEvolution}
%    \partial_t\mathcal{V}=\partial_{\theta}H +i[\mathcal{V},H ],
%\end{equation}
%with initial condition $\mathcal{V}(0)=0$. The previous expression \cref{eq:FisherFormula} is the analytical solution to \cref{eq:VHeisenbergEvolution} in the case where the $H$ is time-independent. Under this evolution, $\mathcal{V}$ remains within the Lie algebra. We can thus express $\mathcal{V}=v_j(t)X_j$, where $v(t)$ are scalar coefficients. In terms of these coefficients, \cref{eq:VHeisenbergEvolution} becomes
%\begin{equation}\label{eq:FisherFormulaTimeDependent}
%    \dot{v}_j(t) = \partial_{\theta}h_j(\theta,t) + i\Gamma_{kl}^jv_k(t)h_l(\theta,t),
%\end{equation}
%with initial condition $v_j(0)=0$. These equations are linear in $v_j(t)$, and we have the same number of differential equations as with the Wei-Norman expansion.
%In contrast the Wei-Norman expansion results in nonlinear differential equations for the coefficients, which suffer from singularities that can prevent even numerical evaluation.

%In the following sections we will apply these results to several problems in metrology. Detailed calculations using both the linear and Wei-Norman parameterisations are included in the supplementary material.

\section{Illustrative examples}\label{sec:IllustrativeExamples}
We will now consider two simple examples which illustrate the power of linear Lie algebra parameterisations. Detailed calculations are provided in the supplementary material.

\subsection{Two-level system}\label{sec:ExampleSpin}
Consider a two-level system with Hamiltonian 
\begin{equation}\label{eq:QubitHamiltonian}
    H=\sin(\theta)\sigma_x+\cos(\theta)\sigma_y,
\end{equation}
where $\theta$ is the parameter we wish to estimate. If we wish to solve this using a Wei-Norman parameterisation 
\begin{equation}\label{eq:UtTwoLevel}
    U(t)=e^{iu_x(t)\sigma_x}e^{iu_y(t)\sigma_y}e^{iu_z(t)\sigma_z},
\end{equation} 
the time-dependent scalar coefficients $u_j(t)$ have equations of motion
\begin{equation}\label{eq:QubitWeiNormanEOM}
    \begin{aligned}
        \dot{u}_x(t) &= -\sin (\theta )-\frac{\cos (\theta ) \sin (2 u_x(t)) \sin (2 u_y(t))}{ \cos (2 u_y(t))}, \\
        \dot{u}_y(t) &=  -\cos (\theta ) \cos (2 u_x(t)), \\
        \dot{u}_z(t) &= -\frac{\cos (\theta ) \sin (2 u_x(t))}{ \cos (2 u_y(t))}.
    \end{aligned}
\end{equation}
These cannot be solved analytically. However, they can be evaluated numerically for a fixed value of $\theta$ \footnote{It may be surprising that \cref{eq:QubitWeiNormanEOM} can be evaluated numerically, given the $\cos\left(2u_y(t)\right))$ terms in the denominator. However, the equations are well-behaved in the vicinity of this point point, hence packages such as Mathematica do not face any issues.}. Since finding the quantum Fisher information using the Wei-Norman method requires differentiating $U(t)$, it too can only be found numerically. The equations of motion obtained from the Wei-Norman parameterisation can differ radically depending on the ordering of the exponentials in \cref{eq:UtTwoLevel}. We have verified however that regardless of operator ordering, the equations of motion can only be evolved numerically, not analytically. 

We can also consider the Wei-Norman expansion in a different Lie algebra basis. The Hamiltonian may be re-written in terms of the operators $\{\sigma_z,\sigma_+,\sigma_-\}$, where $\sigma_{\pm}=(\sigma_x\pm i\sigma_y)/2$ \cite[\S 2.4]{puri_mathematical_2001}. Then parameterising
\begin{equation}\label{eq:TwoLevelUt2}
    U(t)=e^{iu_-(t)\sigma_-}e^{iu_z(t)\sigma_z}e^{iu_+(t)\sigma_+},
\end{equation}
we have equations of motion
\begin{equation}\label{eq:QubitWeiNormanEOMPlusMinus}
    \begin{aligned}
        \dot{u}_-(t)  &= -i\left(e^{-i\theta}-e^{i\theta}u_-(t)^2\right), \\
        \dot{u}_z(t)  &= -e^{i\theta}u_-(t), \\
        \dot{u}_+(t)  &= ie^{i\theta-2iu_z(t)}.
        %\dot{u}_-(t) &= -e^{-i\theta}-e^{i\theta}u_-(t)^2, \\
        %\dot{u}_z(t) &= e^{i\theta}u_-(t), \\
        %\dot{u}_+(t) &= e^{i\theta-2u_z(t)}.
    \end{aligned}
\end{equation}
These equations are singular, and can only be evolved numerically for short times before they blow up. We have verified that this is true regardless of operator ordering in \cref{eq:TwoLevelUt2}. 

However, the equations \cref{eq:QubitWeiNormanEOMPlusMinus} can be solved analytically:
\begin{equation}
    \begin{aligned}
        u_-(t) &= -ie^{-i\theta}\tan(t), \\
        u_z(t) &= -i\log\left(\cos(t)\right), \\
        u_+(t) &= ie^{i\theta}\tan(t).
    \end{aligned}
\end{equation}
The presence of the tangent and log-cosine functions explains why \cref{eq:QubitWeiNormanEOMPlusMinus} cannot be evaluated numerically, since these functions experience periodic singularities. Without the existence of such an analytic solution, which cannot be relied on in general, numerical simulation with the Wei-Norman method would not be possible. 

Our linear methods allow us to bypass the computation of $U(t)$. The formula \cref{eq:FisherGeneratorGeneralForm} gives
\begin{equation}\label{eq:SpinSystemFisherInformation}
    \begin{aligned}
    \mathcal{V}=\sin(t)&\left[\sigma_x\cos(\theta)\cos(t)-\sigma_y\sin(\theta)\cos(t)\right.\\
        &\phantom{[}\left.-\sigma_z\sin(t)\right].
    \end{aligned}
\end{equation}
%where $t$ is the duration of the Hamiltonian interaction. 
The quantum Fisher information assuming an initial state $\rho_0$ is then given by \cref{eq:QFIPure}. However, an advantage of computing $\mathcal{V}$ directly is that we can derive general state-independent results. For example we can see that $\mathcal{V}$ is zero at times $t=2k\pi$ for integer $k$, hence the quantum Fisher information will be zero at these times regardless of the initial quantum state.

To see why this is, we can use the Fernández method to find an exact solution for the dynamics in the Heisenberg picture:
\begin{equation}\label{eq:SpinHeisenbergEOM}
    \begin{aligned}
        \sigma_x(t) &= \left[\cos^2(t)-\sin^2(t)\cos(2\theta)\right]\sigma_x \\
        &\phantom{=}+\sin^2(t)\sin(2\theta)\sigma_y+\sin(2t)\cos(\theta)\sigma_z, \\
        \sigma_y(t) &= \sin^2(t)\sin(2\theta)\sigma_x+\left[\cos^2(t)+\sin^2(t)\cos(2\theta)\right]\sigma_y \\
        &\phantom{=} -\sin(2t)\sin(\theta)\sigma_z, \\
        \sigma_z(t) &= -\sin(2t)\cos(\theta)\sigma_x+\sin(2t)\sin(\theta)\sigma_y \\
        &\phantom{=}+\cos(2t)\sigma_z.
    \end{aligned}
\end{equation}
Linear parameterisations provide an analytical solution valid for all times. Again as this is Heisenberg-picture evolution, any conclusions we draw are independent of the initial quantum state. We can see that at times $t=2k\pi$, the system returns to its initial state regardless of the value of $\theta$. Thus the state at this time cannot provide any information about the parameter.

This example demonstrates that solution with the Wei-Norman method cannot be taken for granted, even for a simple two-level system. One must explore different algebra orderings and bases, each of which requires tedious algebra, to find the parameterisation most amenable to solution. Even then, neither analytical nor numerical solution are guaranteed. In contrast linear parameterisations are independent of operator orderings, and will always provide at least a numerical solution. And as we saw in this case, readily offered a simple analytic solution.

This system has a two-dimensional Hilbert space, thus it could have been simulated exactly without resorting to Lie algebraic methods. However, quantum algorithms apply the Wei-Norman expansion to multi-qubit systems whose Hilbert space grows exponentially \cite{cerezo_variational_2021,goh_lie-algebraic_2023}. In the following section we will consider an infinite-dimensional Hilbert space.

%Since this is a simple two-level system, it is possible that it could be solved in some other way without resorting to Lie algebraic techniques. However, this analysis shows that the standard Wei-Norman expansion and its corresponding nonlinear Lie algebra parameterisation, can fail even for very simple systems.

\subsection{Optical cavity}\label{sec:ExampleCavity}
The Hamiltonian for an optical cavity driven by a coherent field of amplitude $\eta$ with detuning $\Delta$ from the cavity resonance is
\begin{equation}\label{eq:CavityHamiltonian}
    H=\Delta a^{\dagger}a+i\eta\left(a^{\dagger}-a\right).
\end{equation}
The Hilbert space of the optical mode $a$ is infinite-dimensional. Thus simulation using numerical methods in the Fock basis would require truncation, and only be accurate for small times and driving amplitudes. However, the Lie algebra generated by the Hamiltonian is four-dimensional: $\mathcal{L}=\{1,a,a^{\dagger},a^{\dagger}a\}$. This means Lie algebraic methods can provide an exact representation using only four dimensions. 

%The Lie algebra for this system is \emph{solvable}. This means that if we denote by $[\mathcal{L},\mathcal{L}]$ is the set of all commutators of elements in the Lie algebra, the series $\mathcal{L},[\mathcal{L},\mathcal{L}],[\mathcal{L},[\mathcal{L},\mathcal{L}]]$ eventually terminates. 
Let us first suppose that the Hamiltonian is time-independent. Then the Wei-Norman expansion provides an analytical solution for the time-evolution operator:
\begin{equation}\label{eq:UtCavity}
    \begin{aligned}
        U(t) = &\exp\left(\frac{\eta^2}{\Delta^2}\left(1-e^{it\Delta}+it\Delta\right)\right)\exp\left(-i\frac{\eta}{\Delta}\left(1-e^{it\Delta}\right)a\right) \\
            &\times \exp\left(-i\frac{\eta}{\Delta}\left(1-e^{-it\Delta}\right)a^{\dagger}\right)\exp\left(-it\Delta\, a^{\dagger}a\right).
    \end{aligned}
\end{equation}
Suppose we wish to find the photon number at a given time. We must then compute $\mathrm{tr}\{\rho U^{\dagger}(t)a^{\dagger}aU(t)\}$, where $\rho$ is the initial state. Given the complex form of $U(t)$, this could be very tedious \footnote{One could speed up this calculation using the adjoint representation of the Lie algebra, as discussed in the supplementary material.}.

In contrast, the linear parameterisation directly provides the time-evolved operators:
\begin{equation}\label{eq:CavityHeisenbergSolution}
    \begin{aligned}
        a(t) &= e^{-i\Delta t}a-i\frac{\eta}{\Delta}(1-e^{-i\Delta t}), \\
        a^{\dagger}(t) &= e^{i\Delta t}a^{\dagger}+ i\frac{\eta}{\Delta}(1-e^{i\Delta t}), \\
        a^{\dagger}a(t) &=a^{\dagger}a + 4\left(\frac{\eta}{\Delta}\right)^2\sin^2\left(\frac{t\Delta}{2}\right) +\\ &\phantom{=}+i\frac{\eta}{\Delta}\left[(1-e^{it\Delta})a^{\dagger}-(1-e^{-it\Delta })a\right].
    \end{aligned}
\end{equation}
It is now straightforward to compute $\mathrm{tr}\{\rho\,a^{\dagger}a(t)\}$. The linear parameterisation also simplifies many other analyses. For example, Taylor expanding \cref{eq:CavityHeisenbergSolution} around $\Delta=0$ immediately shows that for small detunings, photon number depends quadratically on both time and $\eta$:
\begin{equation}
    \begin{aligned}
    a^{\dagger}a(t) &= t^2\eta^2+t\eta(a+a^{\dagger}) \\
    &\phantom{=}+a^{\dagger}a+i\frac{\eta t^2\Delta}{2}\left(a^{\dagger}-a\right)+\mathcal{O}(\Delta^2).
    \end{aligned}
\end{equation}
It would take a much more involved process to derive the same result from the Wei-Norman solution \cref{eq:UtCavity}.

Let us now consider the quantum Fisher information for a time-dependent Hamiltonian.
%In the case where $\eta$ is constant, both the Wei-Norman method and our \cref{eq:FisherFormula} give
%\begin{equation}
%    \begin{aligned}
%        \mathcal{V} &= 2\frac{\eta^2}{\theta^3}\left[t\theta-\sin(t\theta)\right]-\frac{\eta}{\theta^2}\left[it\theta+(1-e^{it\theta})\right]a \\
%        &\phantom{=}+\frac{\eta}{\theta^2}\left[it\theta-(1-e^{-it\theta})\right]a^{\dagger}+t\,a^{\dagger}a.
%    \end{aligned}
%\end{equation}
%Note that the operators on the right-hand side are the operators at the beginning of the interaction, not the time-evolved Heisenberg picture operators. 
We will suppose that the drive is modulated as $\eta=\eta_0\cos(\omega_dt)$. Both the linear parameterisation and Wei-Norman method can derive an exact formula for $\mathcal{V}$. Since the expression is quite complex, we report only the formula for $\theta=0$:
\begin{equation}
    \begin{aligned}
        \mathcal{V}\rvert_{\theta=0} &= \frac{\eta^2}{4\omega_d^3}\left[2\omega_dt\left(2-\cos(2\omega_dt)\right)\right. \\
        &\hphantom{=\frac{\eta^2}{4\omega_d^3}[[}\left.-8\sin(\omega_dt)+3\sin(2\omega_dt)\right]+ta^{\dagger}a \\
        &\phantom{=} +\frac{\eta_0}{\omega_d^2}\left[1-\omega_dt\sin(\omega_dt)-\cos(\omega_dt)\right](a+a^{\dagger}).
    \end{aligned}
\end{equation}
Taylor expanding this for small times gives:
\begin{equation}
    \mathcal{V}\rvert_{\theta=0}\approx\frac{t^3\eta^2}{3}\left(1+a^{\dagger}a\right)+\frac{t^2\eta}{8}\left(\omega_d^2t^2-4\right)\left(a+a^{\dagger}\right)+\mathcal{O}(t^5).
\end{equation}
This allows us to see that modulating $\eta$ with frequency $\omega_d$ will only affect the QFI if the state has significant variance in the $a+a^{\dagger}$ quadrature. Moreover, this effect is only fourth order in time. 

Exploiting the Lie algebraic structure allowed us to find an exact representation for this infinite-dimensional system, using only four dimensions. Full derivations are provided in the supplementary material. We argue that the linear parameterisation provides a shorter and simpler route to the solution than the Wei-Norman method.

\section{Nonlinear Optomechanics}\label{sec:ExampleOptomechanics}
We will now apply the linear Lie algebra parameterisations to study metrology with nonlinear optomechanics. As before, detailed calculations are provided in the supplementary material. This problem has been extensively studied using the Hamiltonian \cite{qvarfort_time-evolution_2020,schneiter_optimal_2020,qvarfort_constraining_2022}:
\begin{equation}\label{eq:OptomechHamiltonian}
    H=\hbar\omega n+\frac{p^2}{2m}+\frac{1}{2}m\Omega^2x^2-\hbar G n x+mgx.
\end{equation}
Here $n=a^{\dagger}a$, where $a$ is the annihilation operator for an optical field with frequency $\omega$, and $x,p$ are the position and momentum operators of the mechanical oscillator with mass $m$ and frequency $\Omega$. The optomechanical coupling is $G$, and $g$ is gravitational acceleration. Driving terms are neglected because these lead to an infinite-dimensional Lie algebra. However, this Hamiltonian should be approximately valid for high-finesse cavities and short interaction times. Since $n$ communites with teh Hamiltonian, the dynamics will conserve photon number $n$. 

The Hamiltonian generates a ten-dimensional Lie algebra $\mathcal{L}=\left\{1,n,n^2,x,p,x^2,p^2,xp,nx,np\right\}$. In the case where the Hamiltonian is time-independent, we derive using linear parameterisations
\begin{equation}\label{eq:OptomechSolution}
    \begin{aligned}
        x(t)    &= -\frac{2g}{\Omega^2}\sin^2(\Omega t/2)+\frac{2\hbar G}{m\Omega^2}\sin^2(\Omega t/2)n \\
            &\phantom{=}+\cos(\Omega t)x+\frac{1}{m\Omega}\sin(\Omega t)p.
    \end{aligned}
\end{equation}
The momentum $p(t)$ is the time-derivative of this, thus \cref{eq:OptomechSolution} is a fully quantum solution which exactly describes the nonlinear dynamics given by \cref{eq:OptomechHamiltonian}. The solution is also immediately interpretable, in contrast to the Wei-Norman approach which gives $U(t)$ as a product of seven operator exponentials \cite{qvarfort_time-evolution_2020}.

We will follow \cite{schneiter_optimal_2020}, and consider metrology when different terms in \cref{eq:OptomechHamiltonian} oscillate sinusoidally. Suppose the system is subjected to a time-varying gravitational signal with strength $\theta$ and frequency $\omega_g$: 
\begin{equation}\label{eq:GravitationalSignal}
    g(t)=g_0\left(1+\theta\cos(\omega_gt)\right).
\end{equation}
In this scenario we derive the solution
\begin{equation}\label{eq:OptomechgSolution}
        x(t)    = x(t)\rvert_{\theta=0} + \theta g_0\frac{\cos(\Omega t)-\cos(\omega_gt)}{\Omega^2-\omega_g^2},
\end{equation}
where $x(t)\rvert_{\theta=0}$ is the solution \cref{eq:OptomechSolution} to the time-independent Hamiltonian. Thus the effect of the gravitational signal is simply to add a harmonic term proportional to the identity operator. This displays a resonance as $\omega_g\rightarrow\Omega$, and in that limit we find that position grows linearly with time $\sim -(\theta g_0/2\Omega)t\sin(\Omega t)$. 

We emphasise that \cref{eq:OptomechgSolution} is not the expected position, but rather the quantum operator in the Heisenberg picture. Thus we can use this this to understand general features of the dynamics, valid regardless of the initial particular quantum state. For example, \cref{eq:OptomechgSolution} implies that that the effect of a harmonic gravitational signal will always be to add an oscillation to the expected position, while leaving the variance unchanged.

Let us now study the quantum Fisher information for the gravitational signal $\theta$. This will tell us our fundamental sensitivity to gravitational signals of frequency $\omega_g$. Using \cref{eq:TimeDependentFisherGenerator} we derive
\begin{equation}\label{eq:FisherGeneratorGravimetry}
    \begin{aligned}
        \mathcal{V} &= \frac{Gg_0}{\Omega^2\omega_g}\left(1+\frac{\omega_g^2\cos(\Omega t)-\Omega^2\cos(\omega_gt)}{\Omega^2-\omega_g^2}\right)n \\
        &\phantom{=}+  \frac{mg_0\omega_g}{\hbar}\frac{\cos(\omega_gt)-\cos(\Omega t)}{\Omega^2-\omega_g^2}x \\
        &\phantom{=}+ \frac{g_0}{\hbar\Omega}\frac{\omega_g\sin(\Omega t)-\Omega\sin(\omega_gt)}{\Omega^2-\omega_g^2}p.
    \end{aligned}
\end{equation}
We neglect here the component of $\mathcal{V}$ proportional to the identity, which is more complex than the other components. As discussed in \cref{app:Variance}, terms proportional to the identity do not contribute to the quantum Fisher information.

Thus the quantum Fisher information depends only on the modulation frequency $\omega_g$, not the size of the perturbation $\theta$ \footnote{Recall the definition of the quantum Fisher information, as the smallest fluctuation that can be detected about some operating point. We have not found that larger signals ($\lvert\theta\rvert\gg 1$) are equally difficult to detect as small ones ($\lvert\theta\rvert\ll 1$). Rather, fluctuations in larger signals are equally difficult to detect as fluctuations in small signals.}. The quantum Fisher information as a function of $\omega_g$ has previously been computed in \cite{schneiter_optimal_2020} assuming initial coherent optical state and thermal mechanical state, and found to be maximal for constant driving $\Omega=0$ as well as on resonance $\Omega=\omega_g$. From our explicit form, we can see that the QFI depends only on the initial state's covariances of $x,p,n$. At resonance this becomes
\begin{equation}
    \begin{aligned}
        \mathcal{V}\rvert_{\omega_g\rightarrow\Omega} &= -\frac{g_0G}{2\Omega^3}\left[\Omega t\cos(\Omega t)-\sin(\Omega t)\right]n \\
        &\phantom{=}+\frac{mg_0}{2\Omega\hbar}\left[\Omega t\cos(\Omega t)+\sin(\Omega t)\right]x \\
        &\phantom{=}-\frac{g_0}{2\Omega\hbar}t\sin(\Omega t)p.
    \end{aligned}
\end{equation}
At large times, $\mathcal{V}$ grows linearly with the integration time $t$. Its variance will thus grow quadratically. Thus this shows that for any initial quantum state, if the gravitational signal is resonant with the mechanical coupling then the quantum Fisher information will grow quadratically with time.

Let us now suppose that it is the optomechanical coupling $G$ in \cref{eq:OptomechHamiltonian} varies sinusoidally:
\begin{equation}
    G=G_0\left(1+\theta\cos(\omega_Gt)\right).
\end{equation}
This has solution
\begin{equation}\label{eq:OptomechGSolution}
    x(t) = x(t)\rvert_{\theta=0}+\theta\frac{\hbar G_0}{m}\frac{\cos(\omega_Gt)-\cos(\Omega t)}{\Omega^2-\omega_G^2}n.
        %x(t) &= \frac{\hbar G_0}{m\Omega^2\left(\Omega^2-\omega_G^2\right)}\left[2(\Omega^2-\omega_G^2)\sin^2(\omega_Gt)\right. \\
        %    &\phantom{=\hspace{2em}}\left.+\theta\Omega(\Omega-\omega_G)\sin(\Omega t)\right]n \\
        %    &\phantom{=} -\frac{2g}{\Omega^2}\sin^2(\Omega t/2)+\cos(\Omega t)x+\frac{1}{m\Omega}\sin(\Omega t)p.
\end{equation}
Thus a sinusoidal perturbation to coupling frequency adds a harmonic term proportional to the number operator. Again there there is a resonance when $\omega_G=\Omega$, in which case this term grows as $\sim(\theta G_0\hbar/2m\Omega)t\sin(\Omega t)n$.

We also find an explicit form for the quantum Fisher information vector $\mathcal{V}$. Since this is complicated, we will report here only the value at resonance and large times:
\begin{equation}
    \begin{aligned}
        \mathcal{V}\rvert_{\substack{\omega_G\rightarrow\Omega \\ t\rightarrow\infty}} &= -\frac{gG_0}{2\Omega^2}t\Omega\cos(\Omega t)n  \\
        &\phantom{=}+\frac{\hbar G_0^2}{8m\Omega^2}t\left[4 \cos(\Omega t)+\theta\cos(2\Omega t)\right]n^2 \\
        &\phantom{=}-\frac{G_0}{2}t\cos(\Omega t)nx+\frac{G_0}{2m\Omega}t\sin(\Omega t)np.
    \end{aligned}
\end{equation}
Here $\mathcal{V}$ is proportional to the operators $n,n^2,nx,np$ \footnote{As we discuss in the supplementary material, this is exactly the ideal of the Lie algebra generated by the optomechanical coupling term $nx$.}. It is thus the covariances of these quantities in the initial state that determines the quantum Fisher information. Moreover, the quantum Fisher information now has a component depending on the value $\theta$ of the parameter. However, this manifests in the state's variance of $n^2$. We can also see from the form of $\mathcal{V}$ that the quantum Fisher information will again grow linearly with time $t$, regardless of the initial state.

Finally, we consider the case of a harmonically varying mechanical frequency $\Omega$, which to date has only been solved perturbatively. This turns out to be fundamentally different to the previous two cases. Using the linear parameterisation we derive an exact expression for the dynamics in terms of Mathieu functions. Unlike the preceding cases, now all components of $x(t)$ are significantly different from the time-independent solution \cref{eq:OptomechgSolution}. The equations are significantly more complicated, so we provide these in the supplementary material. The quantum Fisher information $\mathcal{V}$ now depends on all components of the Lie algebra, and we find that its calculation requires a perturbative approach.

\section{Discussion}\label{sec:Discussion}

We have shown how the structure of a finite Lie algebra can be leveraged to construct a linear parameterisation of the quantum Fisher information vector. The Wei-Norman expansion describes dynamics on an $n$-dimensional Lie algebra with $n$ nonlinear equations. The Fernández method is able to describe the dynamics by linear equations, at the cost of increasing the dimension to $n^2$. It may have been supposed that this cost must always be paid in order to obtain linear dynamics. However our work shows that this isn't so, and we can in fact find $n$ linear equations which exactly describe the quantum Fisher information. These equations are easier to work with, and are singularity-free. There is likely more work to be done exploring different Lie-algebraic parameterisations of quantum dynamics, and the scenarios in which these are useful.

We note that Wei-Norman solution contains more information than the solution obtained by the linear parameterisations considered here. The Wei-Norman method parameterises the time evolution operator $U(t)$. This can be used to find the quantum Fisher information vector, and the Heisenberg picture dynamics. In addition, $U(t)$ can be used to find the time-evolution of any initial quantum state. Thus linear methods do not remove entirely the need for the Wei-Norman expansion. However, we have shown that linear methods are sufficient to analyse most metrology problems. They can provide a simpler pathway to the solution, and solve problems where the Wei-Norman method fails.

A natural generalisation of our results is to consider multiple parameters \cite{liu_dynamic_2013,sidhu_geometric_2020}. It would also be interesting to explore evolution of the quantum Fisher information vector in non-Markovian systems \cite{qvarfort_enhanced_2023}, and optimal control \cite{pang_optimal_2017}. 
%There are actually several different notions of quantum Fisher information, with their own associated Cramér-Rao bounds \cite{petz_covariance_2001,hayashi_quantum_2017}. These are little researched. Generalise classical measure of distance, this could help compute these.
These results could possibly be applied to Bayesian metrology in a manner analogous to the Bayesian Cramér-Rao bounds \cite{trees_detection_2013}.
It is also necessary to consider open quantum systems and dissipative evolution. Phenomenological models of damping have been considered in \cite{twamley_quantum_1993}. Moreover, the formalism of Liouville space allows Lie algebraic techniques and the Wei-Norman expansion to be applied to open quantum systems \cite{scopa_exact_2019,teuber_solving_2020,qvarfort_master-equation_2021}. It would be fruitful to explore linear parameterisations in this way. However, open quantum systems often lead to infinite Lie algebras. The most natural way to model these is to truncate the Lie algebra at some cutoff. It is possible that the linear parameterisations may be better suited to this truncation than the Wei-Norman expansion.

%One major advantage of our linear parameterisation is that it allows us to study the evolution of the quantum Fisher information vector $\mathcal{V}$ directly. This vector depends only on the form of the interaction, not the quantum state. Studying this evolution may provide an alternate route to understanding metrological problems. For example, an upper bound on the variance of $\mathcal{V}$ provides an upper bound on the quantum Fisher information attainable with any state. The flow equation \cref{eq:VHeisenbergEvolution} is valid whether or not the Lie algebra is finite. Using this we were able to prove that $\mathcal{V}$ evolves in the Lie algebra ideal generated by $\partial_{\theta}H$, allowing us to reduce the dimension of metrology problems. It is likely that further study of this could reveal other facts about the quantum Fisher information.

To date, most application of Lie algebraic techniques to quantum systems have been analytic. However, these methods are also promising for identifying efficient computational bases. For example, the Hilbert space of an optical mode is infinite-dimensional, so exact simulation in Fock space is impossible. Using the Lie algebra however we can identify a four-dimensional basis which allows for exact representation. Numerical simulation in the Wei-Norman basis has been explored for metrology \cite{schneiter_optimal_2020,qvarfort_master-equation_2021,qvarfort_constraining_2022,qvarfort_enhanced_2023}, and quantum computing \cite{goh_lie-algebraic_2023}. It is likely that the non-singular nature of linear parameterisations makes them well-suited to numerical integration of systems with complex time-dependencies.

\section{Acknowledgements}
 Calculations were validated using the Mathematica package \emph{Operator Algebra} co-written with Joseph J Hope, which will be released shortly. We are grateful for discussions with Simon Haine and Lorcán Conlon.

\appendix

\section{Simplifying computation of the variance}\label{app:Variance}
In the case of a pure state, the Fisher information is given by the variance of $\mathcal{V}$, with respect to the initial state at the start of the interaction. Since the tangent vector is an element of the Lie algebra, we will have
\begin{equation}\label{eq:GeneratorExpansion}
    \mathcal{V}=v_j(\theta,t)X_j,
\end{equation}
for some coefficients $v_j$ which depend on both $\theta$ and the interaction time $t$.  
%The Fisher information is the variance of this. This variance may be computed in either the Schr\"odinger or Heisenberg picture.
%The derivation of \cref{eq:FisherGeneratorGeneralForm} depended only on the commutators of the Lie algebra. Both Schr\"odinger and Heisenberg evolution preserve these commutators, thus this expression is valid in both the Schro\"odinger and Heisenberg pictures. We evaluate this using the adjoint representation. This representation is also determined entirely by the commutators, thus does not depend on which picture is used.
%In the Schr\"odinger picture, the variance is computed using the time-dependent state at time $t_0$. In the Heisenberg picture however we use the initial state, but the time-dependent operators at time $t_0$. Thus in the Heisenberg picture the operators in \cref{eq:GeneratorExpansion} are $X_j(t)=u_{jk}(t_0)X_k$, and the Fisher information generator is
%\begin{equation}\label{eq:VarianceTildeSum}
%    \mathcal{V}=\tilde{v}_j(\theta,t_0,T)X_j,
%\end{equation}
%where we have defined
%\begin{equation}
%    \tilde{v}_j(\theta,t_0)=v_k(\theta,T)u_{kj}(t_0).
%\end{equation}
%This expression is valid when the Hamiltonian is time-independent, or if the measurement time $T$ is small compared to the time-dependence of the Hamiltonian.
The variance of \cref{eq:GeneratorExpansion} is defined as
\begin{equation}\label{eq:QFIVariance}
    \begin{aligned}
        \mathrm{var}\left(\mathcal{V}\right) &= \left\langle\mathcal{V}^2\right\rangle-\left\langle\mathcal{V}\right\rangle^2, \\
        %\mathrm{var}\left(\sum_jv_jX_j\right) 
        &= \left\langle \left(\sum_jv_jX_j\right)^2\right\rangle-\left\langle\sum_jv_jX_j\right\rangle^2,
    \end{aligned}
\end{equation}
where $\langle\cdot\rangle$ denotes the expectation value with respect to the initial state, and we have now made sums explicit. Expanding this gives
\begin{equation}
    \begin{aligned}
        &=\sum_{jk}v_jv_k\left\langle X_jX_k\right\rangle-\sum_{jk}v_jv_k\langle X_j\rangle\langle X_k\rangle, \\
        &=\sum_{j}v_j^2\mathrm{var}(X_j)\\
        &\phantom{=}+2\sum_{j<k}v_jv_k\left(\frac{\langle X_jX_k+X_kX_j\rangle}{2}-\langle X_j\rangle\langle X_k\rangle\right).
    \end{aligned}
\end{equation}
Note that the sum $\sum_{j<k}$ is over all distinct pairs $(j,k)$ where $j$ is less than $k$. Let us define the symmetrised covariance between two operators:
\begin{equation}
    \mathrm{covar}(X_j,X_k)=\frac{\langle X_jX_k+X_kX_j\rangle}{2}-\langle X_j\rangle\langle X_k\rangle.
\end{equation}
The variance is then given by
\begin{equation}\label{eq:VarianceSum}
    \mathrm{var}(\mathcal{V})= \sum_jv_j^2\,\mathrm{var}(X_j)+2\sum_{j<k}v_jv_k\,\mathrm{covar}(X_j,X_k).
\end{equation}
Thus to compute the variance \cref{eq:QFIVariance}, we need only compute the variances of the individual operators, and pairwise covariances. Note that the variance of the identity operator is always zero, as is its covariance with any other operator. Thus we can neglect the component of $\mathcal{V}$ which is proportional to the identity.

Finally we note that for a mixed state, the variance of $\mathcal{V}$ gives only an upper bound for the Fisher information. There also exists an exact formula for mixed states \cite{pang_quantum_2014, liu_quantum_2014}.

\clearpage
\newpage

\pagebreak
\begin{widetext}
\begin{center}
  \textbf{\large Supplementary to Quantum metrology with linear Lie algebra parameterisations}
\end{center}

\makeatletter
\setcounter{equation}{0}
\setcounter{figure}{0}
\setcounter{table}{0}
\setcounter{page}{1}
\renewcommand{\theequation}{S\arabic{equation}}
\renewcommand{\thefigure}{S\arabic{figure}}

\vspace{1cm}
\noindent We show in the main body of the text how linear Lie algebra parameterisations can simplify calculations in quantum metrology. This supplementary material contains background information, and detailed examples to illustrate our methods. Ideas from Lie algebras are reviewed in \cref{sec:LieTheory}. As a simple introduction, we then consider a two-level system in \cref{supp:SpinSystem}. The Wei-Norman expansion fails due to singularities, whereas the linear parameterisation can find an exact solution for the dynamics and quantum Fisher information. We next consider a driven cavity in  \cref{app:SingleCavity}. Both the Wei-Norman and linear parameterisations can solve this system exactly, however, we argue that the linear approach is simpler. Finally in \cref{supp:NonlinearOptomech} we consider the model for nonlinear optomechanical dynamics introduced in \cite{schneiter_optimal_2020}. 
The authors find the quantum Fisher information when harmonic modulations are applied to the gravitational field, optomechanical coupling, and mechanical frequency. We consider these same three scenarios, using linear parameterisation. 
We find an solutions for both the dynamics and Fisher information in slightly greater generality than \cite{schneiter_optimal_2020}.
\end{widetext}

\section{Lie algebras and quantum dynamics}\label{sec:LieTheory}
In this section we will give a brief overview of Lie algebras, and summarise the work of Fernández \cite{fernandez_time-evolution_1989} which showed how they can be used to obtain linear parameterisations of dynamics in finite Lie algebras. Lie theory is a broad subject, and we will focus only on the ideas needed for the linear parameterisations. For a pedagogical introduction to Lie algebra for the Wei-Norman expansion we refer to Qvarfort \& Pikovski \cite{qvarfort_solving_2022}.

\subsection{Quick introduction to Lie algebras}
A \emph{Lie algebra} consists of linear combinations of a basis set, where the commutator of any two elements is a linear combination of basis elements.  For example, the set of linear combinations of $\{x,p\}$ is not a Lie algebra, since the commutator $[x,p]$ is $i\hbar$, which cannot be expressed as a linear combination of $x$ and $p$. However, the set of linear combinations of $\{1,x,p\}$ is a Lie algebra. Since there are three elements in the basis, we say that this is three-dimensional. The appearance of commutators is one of the key features of quantum mechanics. It is thus not surprising that Lie algebras can shed light on quantum dynamics. 

We define the \emph{Lie algebra generated by a Hamiltonian} to be the smallest Lie algebra containing all the operators of the Hamiltonian. We can construct this by taking the set of operators in the Hamiltonian and adding to this all operators that can be generated by taking commutators. For example, suppose we are given a Hamiltonian $H=\alpha\sigma_x+\beta\sigma_y$, where $\alpha,\beta$ are real constants and $\sigma_x,\sigma_y$ the Pauli operators. The set of operators in this Hamiltonian is $\{\sigma_x,\sigma_y\}$. Taking the commutator we find $[\sigma_x,\sigma_y]=2i\sigma_z$. Adding this to the set we get $\{\sigma_x,\sigma_y,\sigma_z\}$. This new set is \emph{closed under the commutator}, meaning no new operators can be created by taking commutators. It is therefore a Lie algebra.

As a more complex example, let us consider optomechanics. The standard optomechanical Hamiltonian is
\begin{equation}\label{supp:OMHamiltonian}
    H=\hbar\omega a^{\dagger}a+\frac{p^2}{2m}+\frac{1}{2}m\Omega^2x^2+\hbar Ga^{\dagger}ax+i\eta(a^{\dagger}-a).
\end{equation}
Here $a$ is the cavity annihilation operator and $x,p$ are the position and momentum operators of the mechanical mode. The cavity optical frequency is $\omega$, the mechanical frequency is $\Omega$, and $G$ is the optomechanical coupling. The real parameter $\eta$ describes the strength of the driving laser.

The operators in the Hamiltonian are $\{a,a^{\dagger},a^{\dagger}a,x^2,p^2,a^{\dagger}ax\}$. Since the sum of two operators is also an operator, we have some freedom in how we pick the Lie algebra basis. For example we could take $a^{\dagger}-a$ to be a single operator, then taking commutator with $a^{\dagger}a$ we would then obtain $a^{\dagger}+a$ as another Lie algebra basis element. Which choice of basis is best generally has to be found through experimentation. Note that we must include $a^{\dagger}a$ separately to $a,a^{\dagger}$. This is because inside a Lie algebra we can take sums of operators, multiply them by scalars, and take the commutator of two operators, but we cannot multiply two operators together. Lie algebras are also blind to the Hermitian conjugate operation. Thus through the lens of Lie algebras, the three operators $a$, $a^{\dagger}$, and $a^{\dagger}a$ bear no relation to each other. The same is true for $x,p,x^2,p^2,xp$.
%\footnote{The reader my wonder why we turn to the theory of Lie algebras, if it cannot capture such simple ideas as the product of operators, or the Hermitian conjugate. Surely our lives would be simpler if we used a mathematical structure that did? However, every operation that Lie algebras do not attempt to capture  }

The set of operators in the Hamiltonian is not closed under the commutator. As an example we have $[a,a^{\dagger}ax]=xa$, which we must add to the Lie algebra basis. However, we must then consider the commutator $[xa,a^{\dagger}ax]=x^2a$. This leads to infinite new operators:
\begin{equation}
    [xa,a^{\dagger}ax]=x^2a,\;[x^2a,a^{\dagger}ax]=x^3a,\;[x^3a,a^{\dagger}ax]=x^4a,\ldots
\end{equation}
Thus the basis of the Lie algebra generated by the optomechanical Hamiltonian \cref{supp:OMHamiltonian} must have infinite elements. We say that the Lie algebra is infinite-dimensional. 

The key application of Lie theory in quantum dynamics is that it provides an exact scalar representation of a problem, with dimension equal to the number of elements in the Lie algebra. The theory of infinite-dimensional representations is still being worked out, and is likely a fruitful area for future research. At present however, the application of Lie algebraic methods to quantum dynamics is limited to Hamiltonians which generate finite Lie algebras. For optomechanics, this means dropping the driving term $\eta(a^{\dagger}-a)$ \cite{qvarfort_time-evolution_2020}. The Lie algebra is then ten-dimensional:
\begin{equation}\label{supp:OptomechLieAlgebra}
    \{1,n,n^2,x,p,p^2,x^2,xp,nx,np\}.
\end{equation}
where $n=a^{\dagger}a$ is the photon number operator. The results are then applicable in the regime of highly-reflective mirrors, for short periods of time.

The structure of a Lie algebra comes from its commutation relations, which can be summarised using a \emph{commutator table}. For the Pauli Lie algebra $\{\sigma_x,\sigma_y,\sigma_z\}$ with commutation relation $[\sigma_j,\sigma_k]=2i\epsilon_{jkl}\sigma_l$, the commutator table is:

\begin{center}
    \begin{tabular}{c|ccc}
                    & $\sigma_x$      & $\sigma_y$    & $\sigma_z$ \\
                    \hline
         $\sigma_x$   &               & $2i\sigma_z$  & $-2i\sigma_y$ \\
         $\sigma_y$   & $-2i\sigma_z$   &             & $2i\sigma_x$ \\
         $\sigma_z$   & $2i\sigma_y$   & $-2i\sigma_x$ & \\
    \end{tabular}
\end{center}
The row-$j$ column-$k$ element corresponds to $[\sigma_j,\sigma_k]$. 

To represent the commutator table in a form amenable to algebraic manipulation, we use the \emph{structure constants}, an array of scalars denoted $\Gamma_{jk}^l$. Let $X_j$ represent a general element of the algebra, for example in the Pauli algebra, we have $X_1=\sigma_x$, $X_2=\sigma_y$, $X_3=\sigma_z$. The commutator of any two elements can be written as
\begin{equation}\label{eq:StructureConstants}
    [X_j,X_k]=\Gamma_{jk}^lX_l.
\end{equation}
Here we use the Einstein summation convention of sums whenever there is a repeated index on one side of the equation. In this case $l$ appears in two places on the right-hand side, so we sum over this index. As a concrete example, the relation $[\sigma_x,\sigma_y]=2i\sigma_z$ can be written as:
\begin{equation}
    [X_1,X_2]=0X_1+0X_2+2iX_3,
\end{equation}
which gives
\begin{equation}
    \Gamma_{12}^1=0,\;\Gamma_{12}^2=0,\;\Gamma_{12}^3=2i.
\end{equation}
From the anti-symmetric nature of the commutator $[X_j,X_k]=-[X_k,X_j]$, we have the relation $\Gamma_{jk}^l=-\Gamma_{kj}^l$.

With knowledge of the structure constants, we can commute any commutator between elements of the Lie algebra. Suppose we have some operator $X$ in the Lie algebra. Then we can express it as $A=a_jX_j$, where $a_j$ are scalar coefficients and $\{X_1,\ldots,X_n\}$ form a basis for the algebra. Now let $B=b_kX_k$ be any other operator in the algebra. The commutator of these is:
\begin{equation}
    [A,B] = \left[a_jX_j,b_kX_k\right] = a_jb_k[X_j,X_k] = a_jb_k\Gamma_{jk}^lX_l.
\end{equation}
%Let us briefly summarise. Suppose we have a Hamiltonian, and wish to apply Lie algebraic methods to this.
%\begin{enumerate}
%    \item  The first step is to find the Lie algebra generated by the Hamiltonian. This is done by beginning with the set of all operators in the Hamiltonian, and then enlarging this by all operators you can produce by taking commutators. If the resulting set is infinite, then you must perform an as yet unrealised generalisation of existing literature to infinite dimensions. In many cases however the set will eventually close, and you will have a finite Lie algebra.
%    \item The next step is to 
%\end{enumerate}

\subsection{The adjoint representation}
The commutator $[A,B]$ is a linear function of both arguments $A$ and $B$. This means we should be able to use matrix algebra to simplify its computation. This leads to the \emph{adjoint representation} of the Lie algebra. Suppose we have an $n$-dimensional Lie algebra, with basis $\{X_1,\ldots,X_n\}$. We can represent $X_j$ as a vector in $\mathbb{R}^n$, where the $j$th element is one and all other elements are zero. For example if our algebra is $\{\sigma_x,\sigma_y,\sigma_z\}$, in this representation we have:
\begin{equation}
    \sigma_x=\left(\begin{array}{c}1 \\ 0 \\0 \end{array}\right),\;
    \sigma_y=\left(\begin{array}{c}0 \\ 1 \\0 \end{array}\right),\;
    \sigma_z=\left(\begin{array}{c}0 \\ 0 \\1 \end{array}\right),
\end{equation}
and hence a linear combination is:
\begin{equation}
    a\,\sigma_x+b\,\sigma_y+c\,\sigma_z=\left(\begin{array}{c}a \\ b \\c \end{array}\right).
\end{equation}

Let $X$ be any element of the Lie algebra, and let us consider the map $[X,\cdot]$, which takes an element $A$ and returns $[X,A]$. We will denote this map as $\mathrm{ad}_X$:
\begin{equation}
    \mathrm{ad}_X(A)=[X,A].
\end{equation}
This is a linear map on the Lie algebra. And it follows from linear algebra \cite{axler_linear_2015} that any linear map can be written as a matrix. The $j$th column of the matrix is the column vector corresponding to the action of the map on the $j$-th basis element, i.e. $\mathrm{ad}_X(X_j)$.

Let us look back to the commutator table of the Pauli operators, and use this to construct $\mathrm{ad}_{\sigma_x}$. 
\begin{itemize}
    \item We can see that $\mathrm{ad}_{\sigma_x}(\sigma_x)=0$, hence the first column of $\mathrm{ad}_{\sigma_x}$ is zero. 
    \item $\mathrm{ad}_{\sigma_x}(\sigma_y)=2i\sigma_z$, so the second column is $\left(\begin{array}{ccc} 0 & 0 & 2i\end{array}\right)^T$.
    \item $\mathrm{ad}_{\sigma_x}(\sigma_z)=-2i\sigma_y$, so the second column is $\left(\begin{array}{ccc} 0 &  -2i & 0\end{array}\right)^T$.
\end{itemize}
Thus we find:
\begin{equation}
    \mathrm{ad}_{\sigma_x}=\left(\begin{array}{ccc}0 & 0 & 0 \\ 0 & 0 & -2i \\ 0 & 2i & 0\end{array}\right).
\end{equation}
Thus commutators can now be computed using matrix multiplication:
\begin{equation}
    [\sigma_x,\sigma_y]=\left(\begin{array}{ccc}0 & 0 & 0 \\ 0 & 0 & -2i \\ 0 & 2i & 0\end{array}\right)\left(\begin{array}{c}0 \\ 1 \\0 \end{array}\right)=\left(\begin{array}{c}0 \\ 0 \\2i \end{array}\right)=2i\,\sigma_z.
\end{equation}
Finally, since commutators are linear, we have $[c_kX_k,\cdot]=c_k[X_k,\cdot]$. Thus we have:
\begin{equation}
    \mathrm{ad}_{c_kX_k}=c_k\mathrm{ad}_{X_k},
\end{equation}
which lets us find the adjoint matrix $\mathrm{ad}_X$ for any element of the Lie algebra.

The adjoint representation is especially useful because of a result called the \emph{Hadamard lemma} \cite[\S II.B]{galitski_quantum--classical_2011}:
\begin{equation}\label{eq:HadamardLemma}
    e^{X}Ye^{-X}=e^{\mathrm{ad}_X}Y.
\end{equation}
This can greatly simplify many quantum calculations. In particular, it lets us sidestep infinite-term Baker-Campbell-Hausdorff type expansions. For example, suppose we wish to compute $e^{\theta a^{\dagger}a}i(a^{\dagger}-a)e^{-\theta a^{\dagger}a}$. The set $\{1,a,a^{\dagger},a^{\dagger}a\}$ forms a basis for a four-dimensional Lie algebra. We'll show in \cref{app:SingleCavity} that the adjoint representation gives:
\begin{equation}
    \mathrm{ad}_{a^{\dagger}a}=
    \left(
        \begin{array}{cccc}
               0 & 0 & 0 & 0 
            \\ 0 & -1 & 0 & 0 
            \\ 0 & 0 & 1 & 0 
            \\ 0 & 0 & 0 & 0 
        \end{array}\right).
\end{equation}
The matrix exponential is defined using a power series:
\begin{equation}
    \exp\left(A\right)=\sum_n\frac{A^n}{n!}=1+A+\frac{A^2}{2!}+\cdots.
\end{equation}
This can easily be computed using packages such as Mathematica, which gives:
\begin{equation}
    \exp
    \left(\left( \begin{array}{cccc}
               0 & 0 & 0 & 0 
            \\ 0 & -\theta & 0 & 0 
            \\ 0 & 0 & \theta & 0 
            \\ 0 & 0 & 0 & 0 
    \end{array}\right)\right)
        =
    \left(
        \begin{array}{cccc}
         1 & 0 & 0 & 0 \\
        0 & e^{-\theta } & 0 & 0 \\
        0 & 0 & e^{\theta } & 0 \\
        0 & 0 & 0 & 1 \\
        \end{array} \right).
\end{equation}
Thus we have:
\begin{equation}
    \begin{aligned}
        &e^{\theta a^{\dagger}a}i(a^{\dagger}-a)e^{-\theta a^{\dagger}a}=e^{\theta\,\mathrm{ad}_{a^{\dagger}a}}i(a^{\dagger}-a) \\
        &\phantom{==}=\left(
        \begin{array}{cccc}
         1 & 0 & 0 & 0 \\
        0 & e^{-\theta } & 0 & 0 \\
        0 & 0 & e^{\theta } & 0 \\
        0 & 0 & 0 & 1 \\
\end{array}
\right)\left(\begin{array}{c}0 \\ -i \\ i \\ 0\end{array}\right) \\
    &\phantom{==}=\left( \begin{array}{c} 0 \\ -i e^{-\theta } \\ i e^{\theta } \\ 0 \\
        \end{array} \right) \\
    &\phantom{==}=i(a^{\dagger}e^{\theta}-ae^{-\theta}).
    \end{aligned}
\end{equation}

\subsection{Linear parameterisation for the quantum dynamics}
Here we will describe the method from Fernández \cite{fernandez_time-evolution_1989} for parameterising Heisenberg dynamics that take place on finite Lie algebras. Suppose our Hamiltonian is given by:
\begin{equation}\label{eq:FernandezHam}
    H=h_j(t)X_j,
\end{equation}
where $h_j(t)$ are time-dependent scalar coefficients, and $\{X_j\}$ a basis for a Lie algebra. Note that we are using Einstein summation convention over repeated indices. As before the commutation relations are given by the structure constants
\begin{equation}
    [X_j,X_k]=\Gamma^m_{jk}X_m.
\end{equation}

The key insight is that because Lie algebras are closed under commutators, the time evolution of any operator $X_k$ will remain within the Lie algebra. To see this, the time derivative is
\begin{equation}
    \frac{\mathrm{d}}{\mathrm{d}t}X_k(t)=i[H,X_k(t)].
\end{equation}
Since the Hamiltonian in \cref{eq:FernandezHam} is a sum of Lie algebra basis elements, if $X_k(t)$ is in the Lie algebra, the right-hand side will also be in the Lie algebra. Thus $X_k(t+\mathrm{d}t)$ will be equal to $X_k(t)$ plus an increment which is in the Lie algebra. Since Lie algebras are closed under addition, this means $X_k(t+\mathrm{d}t)$ will be in the Lie algebra.

Let $X_k(t)$ denote the time-evolution of a basis element $X_k$. Since this remains within the Lie algebra, we can expand it as
\begin{equation}
    X_k(t) = u_{km}(t)X_m,
\end{equation}
where $u_{km}$ are time-dependent scalar coefficients. The derivative of this is
\begin{equation}\label{eq:XjDifferentiated}
    \dot{X}_k(t)=\dot{u}_{km}(t)X_m.
\end{equation}
To find the time-evolution of the coefficients, we use the Heisenberg equations of motion
\begin{equation}
    \begin{aligned}
        \dot{X}_k(t) &= i[H(t),X_k(t)],\\
            &= i[h_j(t)X_j(t),X_k(t)].
    \end{aligned}
\end{equation}
Unitary evolution preserves commutators, so we have
\begin{equation}
    [X_j(t),X_k(t)]=\Gamma_{jk}^lX_l(t).
\end{equation}
This gives us
\begin{equation}\label{eq:XJDot}
    \begin{aligned}
        \dot{X}_k(t)    &= ih_j(t)\Gamma_{jk}^lX_l(t), \\
            &= ih_j(t)\Gamma_{jk}^lu_{lm}(t)X_m.
    \end{aligned}
\end{equation}
Substituting \cref{eq:XjDifferentiated} in the left-hand side, equating terms $X_m$, and re-labelling indices, we find the equation of motion for the coefficients:
\begin{equation}
    \dot{u}_{jk}(t)=i\left[h_l(t)\Gamma_{lj}^m\right]u_{mk}(t).
\end{equation}
Defining the matrix:
\begin{equation}\label{eq:FernandezHMatrix}
    \mathbf{H}_{jk}=h_l(t)\Gamma_{lj}^k,
\end{equation}
and letting $\mathbf{u}$ denote the matrix of coefficients $u_{jk}$, we can write the equation of motion in matrix form:
\begin{equation}\label{eq:LieHeisenbergEOM}
    \dot{\mathbf{u}}(t)=i\mathbf{H}\,\mathbf{u}(t).
\end{equation}
The initial condition is $\mathbf{u}(0)$ being the identity matrix:
\begin{equation}
    \mathbf{u}(0)=I.
\end{equation}

Note that in \cref{eq:LieHeisenbergEOM}, $\mathbf{u}$ is a matrix, not a vector. Thus the solution is not provided by eigenvectors and eigenvalues of $\mathbf{H}$. In an $n$-dimensional Lie algebra the dimensions of $\mathbf{u}$ and $\mathbf{H}$ are $n\times n$, meaning the number of equations grow quadratically with the dimension of the Lie algebra. However, $\mathbf{H}$ will be a sparse matrix.

Both this method and the Wei-Norman expansion allow us to convert operator differential equations on finite-dimensional Lie algebras to scalar differential equations. We summarise in the table below the key differences between these approaches:
\begin{center}
    \begin{tabular}{|>{\raggedright\arraybackslash}p{0.25\textwidth}|>{\raggedright\arraybackslash}p{0.25\textwidth}|}
        \hline
        Wei-Norman 
 \cite{wei_global_1964,qvarfort_solving_2022} & Fernández \cite{fernandez_time-evolution_1989} \\
        \hline
        Parameterises the \textbf{time-evolution operator} $U(t)$. & Parameterises \textbf{Heisenberg evolution of the operators} $X_j(t)$. \\
        \hline
        Results in \textbf{$\mathbf{n}$ nonlinear} differential equations. & Results in \textbf{$\mathbf{n^2}$ linear} differential equations. \\
        \hline
        Can have \textbf{singularities}, so only guaranteed to work valid for $t$ in some small interval. & \textbf{Global solution} valid for all $t$. \\
        \hline
    \end{tabular}
\end{center}
In later sections we will provide examples which illustrate these differences.

%\subsection{Summary}
%Here we will try and distil why exactly Lie algebra techniques are useful for quantum dynamics.
%
%Solvable Lie algebra. Not the Pauli matrices.
%We note however that even in solvable Lie algebras there can be singularities which prevent evaluation, as noted in the original paper \cite{wei_global_1964}.
\subsection{Fisher information}
The quantum Fisher information $\mathcal{I}(\theta)$ measures how difficult it is to discriminate between $\rho_{\theta}$ and $\rho_{\theta+\Delta\theta}$, where $\Delta\theta$ is an infinitesimal perturbation. This can be described by an operator $\mathcal{V}_{\theta}$, representing the vector from $\rho_{\theta}$ to $\rho_{\theta+\Delta\theta}$. We will consider unitary evolution $\rho_{\theta}=U(\theta)\rho_0U^{\dagger}(\theta)$, where $U(\theta)$ is some unitary operator, and $\rho_0$ the initial probe state. If $\rho_0$ is a pure state, then the quantum the Fisher information $\mathcal{I}(\theta)$ is given by \cite{liu_quantum_2015}
\begin{equation}\label{eq:QFIFormulaPure}
   \mathcal{I}(\theta)=4\mathrm{var}_{\rho_0}(\mathcal{V}_{\theta}).
\end{equation}
If $\rho_0$ is a mixed state, then \cref{eq:QFIFormulaPure} is an upper bound for the quantum Fisher information. To find the exact value, we must decompose the initial state $\rho_0$ into its eigenbasis $\{|\lambda_n\rangle\}$ \cite{liu_quantum_2015}:
\begin{equation}
    \begin{aligned}
    \mathcal{I}(\theta) &= 4\sum_n\lambda_n\left(\langle\lambda_n|\mathcal{V}_{\theta}^2\lvert\lambda_n\rangle-\langle\lambda_n\rvert\mathcal{V}_{\theta}\lvert\lambda_n\rangle^2\right) \\
    &\phantom{=}-8\sum_{n\ne m}\frac{\lambda_n\lambda_m}{\lambda_n+\lambda_m}\left\lvert\langle\lambda_n\rvert\mathcal{V}_{\theta}\lvert\lambda_m\rangle\right\rvert^2.
    \end{aligned}
\end{equation}
In what follows, we will omit the subscript $\theta$ on $\mathcal{V}_{\theta}$. The utility of the Fisher information comes from the quantum Cram\'er-Rao bound. This states that if we independently measure $N$ copies of $\rho_0$, our error in the parameter is lower bounded by
\begin{equation}
    \mathrm{var}(\theta)\ge\frac{1}{N\mathcal{I}(\theta)}.
\end{equation}

In the main text, we show that for a time-independent Hamiltonian the QFI vector is given by (in units where $\hbar=1$):
\begin{equation}\label{eq:FisherFormula}
    \mathcal{V}=\partial_{\theta}h_j(\theta)\left(\int_0^t\exp\left(-is\,h_k(\theta)\,\mathrm{ad}_{X_k}\right)\mathrm{d}s\right)X_j.
\end{equation}
For a time-dependent Hamiltonian, we show that $\mathcal{V}$ has equation of motion:
\begin{equation}\label{eq:VHeisenbergEvolution}
    \partial_t\mathcal{V}=\partial_{\theta}H +i[\mathcal{V}_{\theta},H].
\end{equation}
If we express the QFI vector in the Lie algebra basis as $\mathcal{V}=v_iX_i$, the coefficients have equation of motion:
\begin{equation}\label{eq:FisherFormulaTimeDependent}
    \dot{v}_j(t) = \partial_{\theta}h_j(\theta,t) + i\Gamma_{kl}^jv_k(t)h_l(\theta,t).
\end{equation}

\subsection{Dimensional reduction using a Lie algebra ideal}\label{sec:Ideals}
We often do not need to use the entire Lie algebra generated by the Hamiltonian. If the dynamics of interest take place on a subset of an algebra, we can use this to reduce the problem dimension. Let the entire Lie algebra $\mathcal{L}$ consist of the operators $\{X_1,\ldots,X_n\}$, and suppose $\mathcal{Y}=\{Y_1,\ldots,Y_m\}$ is a subset of this. If for all $X_j\in\mathcal{L}$ and $Y_k\in\mathcal{Y}$, the commutator $\{X_j,Y_k\}$ is a linear combination of elements of $\mathcal{Y}$, we say that $\mathcal{Y}$ is an \emph{ideal} of the Lie algebra $\mathcal{L}$. If we have some collection of operators, we can generate an ideal by considering repeated commutators with operators in $\mathcal{L}$.

If we want to find the Heiselberg-picture dynamics of some operator using the methods of Fernández \cite{fernandez_time-evolution_1989}, we only need to work on the ideal generated by that operator. This follows from Heisenberg's equation of motion $\dot{Y}_j(t)=i[H(t),Y_j(t)]$. If $Y_j\in\mathcal{Y}$, the right-hand side will consist entirely of elements in the ideal. This reduces the dimension from $n^2$ to $m^2$, which is often substantial.

We can also use an ideal to reduce the dimension of our linear Fisher information vector parameterisations. Let $\mathcal{Y}$ be the ideal generated by all operators $X_j$ in the Hamiltonian where $h_j$ depends on $\theta$. Since $\mathrm{ad}_{X_j}Y_k\in \mathcal{Y}$, the time-independent QFI formula \cref{eq:FisherFormula} will give $\mathcal{V}\in\mathcal{Y}$. The same goes for the time-dependent case \cref{eq:FisherFormulaTimeDependent}, as all terms on the right-hand side will be elements of $\mathcal{Y}$. Thus we can reduce the dimension of the dynamics from $n$ to $m$.

\section{Example: Two-level system}\label{supp:SpinSystem}
As a first example, let us consider a single spin-$1/2$ system. We will consider the Hamiltonian:
\begin{equation}\label{eq:SpinHamiltonian}
    H = \sin(\theta)\sigma_x + \cos(\theta)\sigma_y.
\end{equation}
Here $\sigma_j$ are the Pauli spin operators, and $0\le\theta<2\pi$ is a parameter we are trying to estimate.
%We'll see that this deceptively simple solution cannot be solved even numerically using the Wei-Norman method. The linear Lie algebra parameterizations however will deliver analytic solutions.

\subsection{Pauli Lie algebra}
The Hamiltonian generates a three-dimensional Lie algebra:
\begin{equation}
    \mathcal{L}=\{\sigma_x,\;\sigma_y,\;\sigma_z\}.
\end{equation}
The Pauli matrices satisfy the commutation relation $[\sigma_j,\sigma_k]=i\epsilon_{jkl}\sigma_k$. This gives us the commutation table:
\begin{center}
    \begin{tabular}{c|ccc}
        & $\sigma_x$ & $\sigma_y$ & $\sigma_z$ \\
    \hline
    $\sigma_x$ & $0$ & $2i\,\sigma_z$  & $-2i\,\sigma_y$ \\
    $\sigma_y$ & $-2i\,\sigma_z$ & $0$ & $2i\,\sigma_x$ \\
    $\sigma_z$ & $2i\,\sigma_y$ & $-2i\,\sigma_x$ & $0$ 
    \end{tabular}
\end{center}
From this, we can read out the adjoint representation:
\begin{equation}\label{eq:SpinSystemAdjointRep}
    \begin{gathered}
        \mathrm{ad}_{\sigma_x} =
        \left(\begin{array}{ccc}
            0 & 0 & 0 \\ 
            0 & 0 & -2i \\ 
            0 & 2i & 0 
            \end{array}\right),\; 
        \mathrm{ad}_{\sigma_y} = 
        \left(\begin{array}{ccc} 
            0 & 0 & 2i \\ 
            0 & 0 & 0 \\ 
            -2i & 0 & 0 
        \end{array}\right), \\
        \mathrm{ad}_{\sigma_z} = 
        \left(\begin{array}{ccc} 
            0 & -2i & 0 \\ 
            2i & 0 & 0 \\ 
            0 & 0 & 0 
        \end{array}\right).
    \end{gathered}
\end{equation}

\subsection{Fernández parameterisation of the dynamics}
The Hamiltonian given by \cref{eq:FernandezHMatrix} is:
\begin{equation}
    \mathbf{H}=\left(
        \begin{array}{ccc}
            0 & 0 & -2 i \cos (\theta ) \\
            0 & 0 & 2 i \sin (\theta ) \\
            2 i \cos (\theta ) & -2 i \sin (\theta ) & 0 \\
        \end{array}
    \right).
\end{equation}
The equations of motion \cref{eq:LieHeisenbergEOM} can be solved analytically by Mathematica, which gives:
\begin{equation}\label{eq:SpinHeisenbergEOM}
    \begin{aligned}
        \sigma_x(t) &= \left[\cos^2(t)-\sin^2(t)\cos(2\theta)\right]\sigma_x \\
        &\phantom{=}+\sin^2(t)\sin(2\theta)\sigma_y+\sin(2t)\cos(\theta)\sigma_z, \\
        \sigma_y(t) &= \sin^2(t)\sin(2\theta)\sigma_x+\left[\cos^2(t)+\sin^2(t)\cos(2\theta)\right]\sigma_y \\
        &\phantom{=} -\sin(2t)\sin(\theta)\sigma_z, \\
        \sigma_z(t) &= -\sin(2t)\cos(\theta)\sigma_x+\sin(2t)\sin(\theta)\sigma_y \\
        &\phantom{=}+\cos(2t)\sigma_z.
    \end{aligned}
\end{equation}
We can see that the linear Heisenberg picture provides an analytical solution valid for all times. Note that the time-independent operators on the right-hand side refer to the operators at time $t=0$, prior to Heisenberg evolution. 

\vskip 1.7in
\begin{widetext}
\subsection{Quantum Fisher information}
We will now compute the quantum Fisher information for the parameter $\theta$, using the linear parameterisation developed in this manuscript. Since the Hamiltonian is time-independent, we can use the formula:
\begin{equation}\label{eq:SpinSystemFisherFormula}
    \mathcal{V}=\partial_{\theta}h_j(\theta)\left(\int_0^t\exp\left(-is\,h_k(\theta)\,\mathrm{ad}_{X_k}\right)\mathrm{d}s\right)X_j.
\end{equation}
Let us first consider the argument of the exponential. Using the adjoint representation from \cref{eq:SpinSystemAdjointRep} we find this to be:
\begin{equation}
    -ish_k(\theta)\mathrm{ad}_{X_k}=-is\sin(\theta)\left(\begin{array}{ccc}
            0 & 0 & 0 \\ 
            0 & 0 & -2i \\ 
            0 & 2i & 0 
            \end{array}\right)-is\cos(\theta)\left(\begin{array}{ccc} 
            0 & 0 & 2i \\ 
            0 & 0 & 0 \\ 
            -2i & 0 & 0 
        \end{array}\right)=
        \left(
            \begin{array}{ccc}
             0 & 0 & 2 s \cos (\theta ) \\
             0 & 0 & -2 s \sin (\theta ) \\
             -2 s \cos (\theta ) & 2 s \sin (\theta ) & 0 \\
            \end{array}
        \right).
\end{equation}
Computing the matrix exponential (in Mathematica) we find:
\begin{equation}
    \exp\left(-is\,h_k(\theta)\,\mathrm{ad}_{X_k}\right)=
    \left(
        \begin{array}{ccc}
         \cos ^2(s)-\cos (2 \theta ) \sin ^2(s) & \sin (2 \theta ) \sin ^2(s) & \cos (\theta ) \sin (2 s) \\
         \sin (2 \theta ) \sin ^2(s) & \cos (2 \theta ) \sin ^2(s)+\cos ^2(s) & -\sin (\theta ) \sin (2 s) \\
         -\cos (\theta ) \sin (2 s) & \sin (\theta ) \sin (2 s) & \cos (2 s) \\
        \end{array}
    \right).
\end{equation}
Integrating this from $s=0$ to $s=t$ gives:
\begin{equation}
    \begin{aligned}
        &\int_0^t\exp\left(-is\,h_k(\theta)\,\mathrm{ad}_{X_k}\right)\mathrm{d}s=\\
        &\phantom{=}=\left(
            \begin{array}{ccc}
             \frac{1}{4} \left[\cos (2 \theta ) (\sin (2 t)-2 t)+2 t+\sin (2 t)\right] & \sin (\theta ) \cos (\theta ) \left[t-\sin (t) \cos (t)\right] & \cos (\theta ) \sin ^2(t) \\
             \sin (\theta ) \cos (\theta ) \left[t-\sin (t) \cos (t)\right] & t \cos ^2(\theta )+\sin ^2(\theta ) \sin (t) \cos (t) & -\sin (\theta ) \sin ^2(t) \\
             -\cos (\theta ) \sin ^2(t) & \sin (\theta ) \sin ^2(t) & \sin (t) \cos (t) \\
            \end{array}
        \right).
    \end{aligned}
\end{equation}
\end{widetext}

In the adjoint representation, we have 
\begin{equation}
    \sigma_x\sim\left(\begin{array}{c}1 \\ 0 \\ 0\end{array}\right),\;
    \sigma_y\sim\left(\begin{array}{c}0 \\ 1 \\ 0\end{array}\right),\;
    \sigma_z\sim\left(\begin{array}{c}0 \\ 0 \\ 1\end{array}\right).
\end{equation}
Thus using the above and \cref{eq:SpinSystemFisherFormula} we have
\begin{equation}
    \mathcal{V}=
        \left(
            \begin{array}{c}
                \cos(\theta)\sin(t)\cos(t) \\
                -\sin(\theta)\sin(t)\cos(t) \\
                -\sin(t)^2
            \end{array}
        \right),
\end{equation}
which corresponds to the operator
\begin{equation}\label{eq:SpinSystemFisherInformation}
    \mathcal{V}=\sin(t)\left[\sigma_x\cos(\theta)\cos(t)-\sigma_y\sin(\theta)\cos(t)-\sigma_z\sin(t)\right].
\end{equation}
Thus we have a simple analytical form for $\mathcal{V}$, which is valid for all times $t$. 

An immediate consequence of \cref{eq:SpinSystemFisherInformation} is that the quantum Fisher information is zero at period intervals $t=2k\pi$ for integer $k$. If we return to the dynamics \cref{eq:SpinHeisenbergEOM}, we see that at times $t=2k\pi$ the operators are $\sigma_x(t)=\sigma_x$, $\sigma_y(t)=\sigma_y$, $\sigma_z(t)=\sigma_z$, irrespective of the value of $\theta$.

Note that it was not necessary to find the analytical solution \cref{eq:SpinHeisenbergEOM} in order to derive the Fisher information. However, these two calculations show that systems can be solved entirely using linear Lie algebra parameterisations. In the next section we will compare this with the (nonlinear) Wei-Norman approach.

\subsection{Comparison with the Wei-Norman parameterisation}
In the Wei-Norman expansion, we parameterise the time-evolution operator as a product of exponents of the Lie algebra basis: \cite{qvarfort_solving_2022}
\begin{equation}\label{eq:WNexponent}
    U(t)=e^{iu_x(t)\sigma_x}e^{iu_y(t)\sigma_y}e^{iu_z(t)\sigma_z}.
\end{equation}
The initial condition $U(0)=I$ gives $u_j(t)=0$. Differentiating \cref{eq:WNexponent} and using the Schrodinger equation $\frac{dU}{dt}U^{-1}=-iH$ we get:

\begin{equation}\label{eq:cat}
    \begin{aligned}
        H &= i\frac{dU}{dt}U^{-1}, \\
          &= -\dot u_x(t)\sigma_x-
          U_x(t)\dot u_y(t)\sigma_yU_x^{-1}(t)-\\
          &-U_x(t)U_y(t)\dot u_z(t)\sigma_zU_y^{-1}(t)U_x^{-1}(t).
    \end{aligned}
\end{equation}

Next, we compute sandwich terms $U_j(t)XU_j^{-1}(t)$. This is often performed using the Baker-Campbell-Haussdorff formula, which results in much algebra. We note that this could also be computed more efficiently using the adjoint representation and the Hadamard lemma. Either method gives:
\begin{equation}
    \begin{aligned}
        &U_x(t)\sigma_yU_x^{-1}(t) \\
        &\hspace{1em}=-\sigma_z\sin(2u_x(t))+\sigma_y\cos(2u_x(t)),\\
        &U_x(t)U_y(t)\sigma_zU_y^{-1}(t)U_x^{-1}(t)  \\
        &\hspace{1em}=\cos(2u_y(t))(\sigma_z\cos(2u_x(t))+
        \sigma_y\sin(2u_x(t))) \\
        &\hspace{1em}\phantom{=}-\sigma_x\sin(2u_y(t)).
    \end{aligned}
\end{equation}

We then substitute these into \cref{eq:cat}. Since the Lie algebra basis operators $\sigma_j$ are linearly independent, we can equate coefficients of the $\sigma_j$. From this we derive a set of differential equations for functions $u_j(t)$, which depend on the original coefficients Hamiltonian $H=\sin(\theta)\sigma_x+\cos(\theta)\sigma_y$:
\begin{equation}
    \begin{aligned}
        \sigma_x &:\; -\dot u_x(t)+\dot u_z(t)\sin(2u_y(t)) \\
        &\hspace{2em}=\sin\theta,\\
        \sigma_y &:\; -\dot u_y(t)\cos(2u_x(t))-\dot u_z(t)\cos(2u_y(t))\sin(2u_x(t)) \\
        &\hspace{2em}=\cos\theta,\\
        \sigma_z &:\; \dot u_y(t)\sin(2u_x(t))-\dot u_z(t)\cos(2u_y(t))\cos(2u_x(t)) \\
        &\hspace{2em}=0.
    \end{aligned}
\end{equation}
This system can be re-arranged as
\begin{equation}
    \begin{aligned}
        \dot{u}_x(t) &= -\sin (\theta )-\frac{\cos (\theta ) \sin (2 u_x(t)) \sin (2 u_y(t))}{ \cos (2 u_y(t))}, \\
        \dot{u}_y(t) &=  -\cos (\theta ) \cos (2 u_x(t)), \\
        \dot{u}_z(t) &= -\frac{\cos (\theta ) \sin (2 u_x(t))}{ \cos (2 u_y(t))}.
    \end{aligned}
\end{equation}
% \begin{equation}
%     \begin{aligned}
%         \dot{u}_x(t) &=-i \sin (\theta )+ \frac{i \cos (\theta ) \sinh (2 u_x(t)) \sinh (2 u_y(t))}{\cosh ^2(2 u_x(t)) \cosh (2 u_y(t))-\sinh ^2(2 u_x(t)) \cosh (2 u_y(t))}, \\
%         \dot{u}_y(t) &= -\frac{i \cos (\theta ) \cosh (2 u_x(t)) \cosh (2 u_y(t))}{\cosh ^2(2 u_x(t)) \cosh (2 u_y(t))-\sinh ^2(2 u_x(t)) \cosh (2 u_y(t))}, \\
%         \dot{u}_z(t) &= -\frac{\cos (\theta ) \sinh (2 u_x(t))}{\cosh ^2(2 u_x(t)) \cosh (2 u_y(t))-\sinh ^2(2 u_x(t)) \cosh (2 u_z(t))}.
%     \end{aligned}
% \end{equation}

Unlike the equations provided by the linear parameterisation, these do not have an analytical solution. However, these can be numerically evaluated. 

We can also try a different basis for the Lie algebra. If we define \cite{puri_mathematical_2001}
\begin{equation}
    \sigma_{\pm}=\frac{\sigma_x\pm i\sigma_y}{2},
\end{equation}
an equivalent basis for the Lie algebra is 
\begin{equation}
    \mathcal{L}'=\{\sigma_z,\sigma_+,\sigma_-\}.
\end{equation}
This has commutator table :
\begin{center}
    \begin{tabular}{c|ccc}
        & $\sigma_z$ & $\sigma_+$ & $\sigma_-$ \\
    \hline
    $\sigma_z$ & $0$ & $2\sigma_+$  & $-2\sigma_-$ \\
    $\sigma_+$ & $-2\sigma_+$ & $0$ & $\sigma_z$ \\
    $\sigma_-$ & $2\sigma_-$ & $-\sigma_z$ & $0$ 
    \end{tabular}
\end{center}
In this basis the Hamiltonian becomes
\begin{equation}
    H=i\left(-\sigma_+e^{i\theta}+\sigma_-e^{-i\theta}\right),
\end{equation}
which is Hermitian since $\sigma_+^{\dagger}=\sigma_-$.

We then express
\begin{equation}
    %U(t)=e^{u_z(t)\sigma_z}e^{u_+(t)\sigma_+}e^{u_-(t)\sigma_-},
    U(t)=e^{iu_-(t)\sigma_-}e^{iu_z(t)\sigma_z}e^{iu_+(t)\sigma_+},
\end{equation}
and follow the same procedure to derive equations of motion for $u_z,u_+,u_-$. This gives us equations:
%\begin{equation}
   % \begin{aligned}
   %     \dot{u}_z(t) &= -e^{-i\theta+u_z(t)}u_+(t), \\
   %     \dot{u}_+(t) &= -e^{i\theta-u_z(t)}+\frac{1}{2}e^{i\theta+u_z(t)}u_+(t)^2, \\
   %     \dot{u}_-(t) &= e^{-i\theta+u_z(t)},
  %  \end{aligned}
%\end{equation}
%
%\begin{equation}
%    \begin{aligned}
%        \dot{u}_z(t) &= e^{2u_z(t)}\dot{u}_+(t)u_-(t), \\
%        \dot{u}_+(t) &= e^{i\theta-2u_z(t)}, \\
%        \dot{u}_-(t) &= -e^{-i\theta}+\dot{u}_+(t)u^2_-(t)e^{2u_z(t)}-2\dot{u}_z(t)u_-(t),
%    \end{aligned}
%\end{equation}
%which simplify to
\begin{equation}\label{eq:QubitWeiNormanEOMPlusMinus}
    \begin{aligned}
        \dot{u}_-(t) &= -ie^{-i\theta}+ie^{i\theta}u_-(t)^2, \\
        \dot{u}_z(t) &= -e^{i\theta}u_-(t), \\
        \dot{u}_+(t) &= ie^{i\theta-2iu_z(t)}.
    \end{aligned}
\end{equation}
Numerical evolution shows that these equations are singular, and `blow up' after short times. However, we can find an analytical solution: 

\begin{equation}
    \begin{aligned}
        u_-(t) &= -ie^{-i\theta}\tan(t), \\
        u_z(t) &= -i\log\left(\cos(t)\right), \\
        u_+(t) &= ie^{i\theta}\tan(t).
    \end{aligned}
\end{equation}

In a more complex case with time-dependent $\theta$ however neither analytic nor numeric solution would be possible.

Thus we can see that success of the Wei-Norman expansion depends on the basis of the Lie algebra, and in fact also the order of exponents in the product \cref{eq:WNexponent}. These correspond to different choices of coordinate charts on the space of quantum dynamics. There is no way of knowing beforehand which order works the best, and if there will be an ordering which is singularity-free. Thus finding the optimal choice of basis, and ordering of operators in the exponential, can be a very time-consuming task involving a lot of algebra. The linear parameterisation in contrast is independent of ordering, and since it is singularity-free it is always solvable at least numerically.

\section{Example: Optical cavity}\label{app:SingleCavity}
Let us now consider an optical cavity. This has Hamiltonian 
\begin{equation}\label{eq:SingleCavityHamiltonian}
    H=\Delta a^{\dagger}a+i\eta\left(a^{\dagger}-a\right),
\end{equation}
where $\Delta$ is the detuning between the driving frequency and cavity resonance, and $\eta$ the amplitude of the coherent driving field. We will consider both constant and time-dependent driving $\eta$, and assume unitary dynamics (i.e. without loss).

\subsection{Lie algebra for the single cavity}\label{app:CavityAlgebraDynamics}
The Hamiltonian \cref{eq:SingleCavityHamiltonian} generates a four-dimensional Lie algebra:
\begin{equation}
    \mathcal{L}=\{1,a,a^{\dagger},a^{\dagger}a\}.
\end{equation}
This has commutator table:
\begin{center}
    \begin{tabular}{c|cccc}
        & $1$ & $a$ & $a^{\dagger}$ & $a^{\dagger}a$ \\
    \hline
    $1$ & $0$ & $0$  & $0$ & $0$ \\
    $a$ & $0$ & $0$  & $1$ & $a$ \\
    $a^{\dagger}$ & $0$ & $-1$ & $0$ & $-a^{\dagger}$ \\
    $a^{\dagger}a$ & $0$ & $-a$ & $a^{\dagger}$ & $0$
    \end{tabular}
\end{center}

%The element in the $j$th row and $k$th column corresponds to the commutator $[X_j,X_k]$. 
By comparing with \cref{eq:StructureConstants} we can compute the algebra's structure constants $\Gamma_{jk}^m$.
Anti-symmetry of the commutator $[X_j,X_k]=[X_k,X_j]$ implies $\Gamma_{jk}^m=-\Gamma_{kj}^m$, and $\Gamma_{jj}^m=0$. Since $1$ commutes with everything we have $\Gamma_{1j}^m=0$. The other values are
\begin{equation}
    \begin{aligned}
        \Gamma_{23}^m &= \left(\begin{array}{cccc} 1 & 0 & 0 & 0 \end{array}\right), \\
        \Gamma_{24}^m &= \left(\begin{array}{cccc} 0 & 1 & 0 & 0 \end{array}\right), \\
        \Gamma_{34}^m &= \left(\begin{array}{cccc} 0 & 0 & -1 & 0\end{array}\right).
    \end{aligned}
\end{equation}

The commutator table also gives us the adjoint representation:
\begin{equation}
    \begin{gathered}
    \mathrm{ad}_1=\left(
        \begin{array}{cccc}
               0 & 0 & 0 & 0 
            \\ 0 & 0 & 0 & 0 
            \\ 0 & 0 & 0 & 0 
            \\ 0 & 0 & 0 & 0 
        \end{array}\right), \;
    \mathrm{ad}_a=\left(
    \begin{array}{cccc}
           0 & 0 & 1 & 0 
        \\ 0 & 0 & 0 & 1 
        \\ 0 & 0 & 0 & 0 
        \\ 0 & 0 & 0 & 0 
    \end{array}\right),  \\
    \mathrm{ad}_{a^{\dagger}}=\left(
        \begin{array}{cccc}
               0 & -1 & 0 & 0 
            \\ 0 & 0 & 0 & 0 
            \\ 0 & 0 & 0 & -1 
            \\ 0 & 0 & 0 & 0 
        \end{array}\right), \;
    \mathrm{ad}_{a^{\dagger}a}=\left(
        \begin{array}{cccc}
               0 & 0 & 0 & 0 
            \\ 0 & -1 & 0 & 0 
            \\ 0 & 0 & 1 & 0 
            \\ 0 & 0 & 0 & 0 
        \end{array}\right).
    \end{gathered}
\end{equation}

\subsection{Fernández parameterisation of the dynamics}
%We will now compute the Heisenberg dynamics using the method of \cite{fernandez_time-evolution_1989}. 
The Hamiltonian coefficients $h_l$ are:
\begin{equation}
    h_1=0,\; h_2=-i\eta,\;h_3=i\eta,\;h_4=\Delta.
\end{equation}
The Hamiltonian matrix \cref{eq:FernandezHMatrix} is then
\begin{equation}
    \mathbf{H} = \left(\begin{array}{cccc}
        0 & 0 & 0 & 0 \\
        -i\eta & -\Delta & 0 & 0 \\
        -i\eta & 0 & \Delta & 0 \\
        0 & -i\eta & -i\eta & 0
    \end{array}\right).
\end{equation}

We will first suppose that $\eta$ is independent of time. Solving the equation of motion \cref{eq:LieHeisenbergEOM} we find
\begin{equation}
    u=\left(\begin{array}{cccc}
        1 & 0 & 0 & 0 \\
        -i\frac{\eta}{\Delta}(1-e^{-i\Delta t}) & e^{-i\Delta t} & 0 & 0 \\
        i\frac{\eta}{\Delta}(1-e^{i\Delta t}) & 0 & e^{i\Delta t} & 0 \\
        4\left(\frac{\eta}{\Delta}\right)^2\sin^2\left(\frac{\Delta t}{2}\right) & -i\frac{\eta}{\Delta}(1-e^{-i\Delta t}) & i\frac{\eta}{\Delta}(1-e^{i\Delta t}) & 1
    \end{array}\right).
\end{equation}

The rows of this give us the Heisenberg evolution of the individual operators:
\begin{equation}\label{eq:CavityHeisenbergSolution}
    \begin{aligned}
        a(t) &= e^{-i\Delta t}a-i\frac{\eta}{\Delta}(1-e^{-i\Delta t}), \\
        a^{\dagger}(t) &= e^{i\Delta t}a^{\dagger}+ i\frac{\eta}{\Delta}(1-e^{i\Delta t}), \\
        a^{\dagger}a(t) &=a^{\dagger}a + 4\left(\frac{\eta}{\Delta}\right)^2\sin^2\left(\frac{t\Delta}{2}\right) +\\ &\phantom{=}+i\frac{\eta}{\Delta}\left[(1-e^{it\Delta})a^{\dagger}-(1-e^{-it\Delta })a\right].
    \end{aligned}
\end{equation}
Note that we have omitted the trivial evolution of the Lie algebra element $1(t)=1$, and operators in \cref{eq:CavityHeisenbergSolution} which are not proportional to $a,a^{\dagger},a^{\dagger}a$ are proportional to the identity. Despite factors of $\Delta$ in the denominator, these equations are not singular in the limit $\Delta\rightarrow0$, which can be confirmed by Taylor expanding them to first order in $\Delta$.

The equations \cref{eq:CavityHeisenbergSolution} are exact, regardless of how many photons are in the cavity. The Hilbert space corresponding to the optical mode is infinite-dimensional, but we can obtain an exact representation using only a four-dimensional Lie algebra. This is because the dynamics of the system usually occupy a low-dimensional submanifold of the full Hilbert space. The Lie algebra approach allows us to find a natural parameterisation of this submanifold.

Let us examine this evolution in various regimes.
\begin{description}
    \item[No driving] If we set $\eta=0$, then the number of photons in the cavity becomes $a^{\dagger}a(t)=a^{\dagger}a$. Then photon number is conserved, as should be expected in the case of a closed system with no damping. The annihilation operator becomes $a(t)=e^{-i\Delta t}a$, so the coherent state amplitude will simply oscillate in phase space.
    \item[Driving on resonance] Taylor expanding about $\Delta=0$ gives:
    \begin{equation}
        \begin{aligned}
        a^{\dagger}a(t) &= t^2\eta^2+t\eta(a+a^{\dagger}) \\
        &\phantom{=}+a^{\dagger}a+i\frac{\eta t^2\Delta}{2}\left(a^{\dagger}-a\right)+\mathcal{O}(\Delta^2).
        \end{aligned}
    \end{equation}
    Thus when driven on resonance the photon number increases quadratically with time. The coherent state amplitude is
    \begin{equation}
        a(t)=(a+t\eta)-i\frac{ t^2\Delta}{2}\left(a^{\dagger}-a\right)+\mathcal{O}(\Delta^2),
    \end{equation}
    which to first order grows linearly with time.
\end{description}
%{\color{red} JT: I am wondering about these equations - is it true that in the Heisenberg picture $\langle X(t)\rangle=Tr[X(t)\rho_0]$, where $\rho_0$ is the initial state of the system and $X(t)$, is the time dependent operator in the Heisenberg picture? If so, let us consider the case with $\eta=0$, then in that case the Hamiltonian is that of a harmonic oscillator with frequency $\omega$, and if we start with a coherent state$\rho_0=|\alpha\rangle\langle \alpha |$, then in Eq. (\ref{eq:15}), $N(t)$ is a constant and $A(t)$, and $D(t)$ rotate phase. This is correct. Now look at the case $\omega=0$. In that case the Hamiltonian is $H=\eta p$, and the action of this Hamiltonian on the coherent state is to displace it along the $x$-axis $\alpha\rightarrow \alpha+\eta t$, and the coherent state's energy grows unbounded. However in the Eq. (\ref{eq:15}), it is not apparaent to me that $Tr[N(t)\rho_0]$, grows with time, indeed for an intial coherent state with  $\alpha\in {\mathbb R}$, $N(t)$ just seems to oscillate. So I suspect there are some secular terms in the solution which have been perhaps missed?}

\subsection{Quantum Fisher information for time-independent Hamiltonian}
Let us take the cavity detuning as the parameter we wish to estimate:
\begin{equation}
    \theta=\Delta.
\end{equation}
The Fisher information vector is given by:
\begin{equation}
    \mathcal{V}=\partial_{\theta}h_j(\theta)\left(\int_0^t\exp\left(-is\,h_k(\theta)\,\mathrm{ad}_{X_k}\right)\mathrm{d}s\right)X_j.
\end{equation}
In our case only $h_4$ depends on $\theta$, and $\partial_{\theta}h_4=1$. Thus the vector is
\begin{equation}
    \mathcal{V} = \int_0^te^{-is\,\left(-i\eta\mathrm{ad}_A+i\eta\mathrm{ad}_D+\theta\mathrm{ad}_N\right)}N\,\mathrm{d}s.
\end{equation}

Using the adjoint representation computed earlier we can evaluate:
\begin{equation}
    \begin{aligned}
    &e^{-is\left(\theta\mathrm{ad}_{a^{\dagger}a}+i\eta\mathrm{ad}_{a^{\dagger}}-i\eta\mathrm{ad}_{a}\right)} \\
    &=\left(
    \begin{array}{cccc}
    1 & -i\frac{\eta}{\theta}\left(1-e^{is\theta}\right) & i\frac{\eta}{\theta}\left(1-e^{-is\theta}\right) & 2\left(\frac{\eta}{\theta}\right)^2(1-\cos(s\theta)) \\
    0 & e^{is\theta} & 0 & -i\frac{\eta}{\theta}\left(1-e^{is\theta}\right) \\
    0 & 0 & e^{-is\theta} & i\frac{\eta}{\theta}(1-e^{-is\theta})
    \\ 0 & 0 & 0 & 1
    \end{array}
    \right).
    \end{aligned}
\end{equation}
Note that though there are factors of $\theta$ in the denominator, Taylor expanding the sines, cosines, and exponentials shows that all quantities have well-defined limits as $\theta\rightarrow 0$. Multiplying this by $N=(0,0,0,1)$ gives the integrand:
\begin{equation}
    \int_0^t\mathrm{d}s\,\left(\begin{array}{c}
    2\left(\frac{\eta}{\theta}\right)^2(1-\cos(s\theta)) \\
    -i\frac{\eta}{\theta}\left(1-e^{is\theta}\right)  \\
    i\frac{\eta}{\theta}(1-e^{-is\theta}) \\
    1
    \end{array}
        \right)
        =\left(\begin{array}{c}
            2\frac{\eta^2}{\theta^3}\left[t\theta-\sin(t\theta)\right] \\
            -\frac{\eta}{\theta^2}\left[it\theta+(1-e^{it\theta})\right] \\
            \frac{\eta}{\theta^2}\left[it\theta-(1-e^{-it\theta})\right] \\
            t
        \end{array}\right)
\end{equation}

Putting all this together, we find
\begin{equation}\label{eq:SingleCavityGenerator}
    \begin{aligned}
        \mathcal{V}&= 2\frac{\eta^2}{\theta^3}\left[t\theta-\sin(t\theta)\right]-\frac{\eta}{\theta^2}\left[it\theta+(1-e^{it\theta})\right]a \\
        &\phantom{=}+\frac{\eta}{\theta^2}\left[it\theta-(1-e^{-it\theta})\right]a^{\dagger}+t\,a^{\dagger}a.
    \end{aligned}
\end{equation}

Evaluating this at $\theta=0$, corresponding to the cavity being driven at resonance, gives
\begin{equation}
    \left.\mathcal{V}\right\rvert_{\theta=0} = \frac{t^3\eta^2}{3}-\frac{t^2\eta}{2}\left(a+a^{\dagger}\right)+ta^{\dagger}a.
\end{equation}
From this we can see how our ability to estimate the cavity detuning depends on the driving strength, evolution time, and initial cavity state. For example in the case of strong driving, our sensitivity grows with the cube of time.

Note that we can also find $\mathcal{V}$ using our formula \cref{eq:FisherFormulaTimeDependent} derived for a time-dependent Hamiltonian:
\begin{equation}
    \dot{v}(t)=\partial_{\theta}h_j(\theta,t)+i\Gamma_{kl}^jv_k(t)h_l(\theta,t).
\end{equation}
As before we have
\begin{equation}
    h_1=0,\;h_2=-i\eta,\;h_3=i\eta,\;h_4=\theta,
\end{equation}
where $\eta$ is time-independent. The Fisher information generator is
\begin{equation}
    \mathcal{V}=v_j(t)X_j,
\end{equation}
where the $v_j$ have equations of motion
\begin{equation}
    \begin{aligned}
        v_1'(t) &= -\eta\,v_2(t)-\eta\,v_3(t), \\
        v_2'(t) &= i\theta\,v_2(t)-\eta\,v_4(t), \\
        v_3'(t) &= -i\theta\,v_3(t)-\eta\,v_4(t), \\
        v_4'(t) &= 1.
    \end{aligned}
\end{equation}
Solving this differential equation with the initial condition $v_j(0)=0$ yields exactly \cref{eq:SingleCavityGenerator}.

To find the quantum Fisher information, we compute the covariance of $\mathcal{V}$ with respect to the initial state as in \cref{eq:QFIFormulaPure}. We will now do this for two cases: Fock and coherent states.

\subsubsection{Fock state}
Let's suppose our system begins in a Fock state $|n\rangle$. The expectation values of Lie algebra operators in \cref{eq:SingleCavityGenerator} are
\begin{equation}
    \begin{aligned}
        \langle 1\rangle &= 1, \\
        \langle a+a^{\dagger}\rangle &= 0, \\
        \langle a^{\dagger}a\rangle &= n.
    \end{aligned}
\end{equation}
We now must also compute the squared terms:
\begin{equation}
    \begin{aligned}
        \left\langle (a+a^{\dagger})^2\right\rangle &= \left\langle a^2+aa^{\dagger}+a^{\dagger}a+(a^{\dagger})^2\right\rangle, \\
            &= (n+1)+n, \\
            &= 2n+1. \\
        \left\langle (a^{\dagger}a)^2\right\rangle &= n^2.
    \end{aligned}
\end{equation}
Finally we have the cross terms
\begin{equation}
    \begin{aligned}
        \left\langle (a+a^{\dagger})(a^{\dagger}a)\right\rangle &= \left\langle aa^{\dagger}a+(a^{\dagger})^2a \right\rangle, \\
        &= 0, \\
        &= \left\langle (a^{\dagger}a)(a+a^{\dagger})\right\rangle.
    \end{aligned}
\end{equation}

The covariance of any operator with itself is the variance of that operator. From the above, the variances of the operators in $\mathcal{V}$ are
\begin{equation}
    \begin{aligned}
        \mathrm{var}(1) &= \langle 1^2\rangle-\langle 1\rangle^2=0, \\
        \mathrm{var}(a+a^{\dagger}) &= \left\langle (a+a^{\dagger})^2\right\rangle-\left\langle a+a^{\dagger}\right\rangle^2 = 2n+1, \\
        \mathrm{var}(a^{\dagger}a) &= \left\langle (a^{\dagger}a)^2\right\rangle-\left\langle a^{\dagger}a\right\rangle^2 = 0.
    \end{aligned}
\end{equation}
The covariance of the identity with any operator is zero. This leaves
\begin{equation}
    \begin{aligned}
        \mathrm{covar}\left(a^{\dagger}a , a+a^{\dagger}\right) &= \frac{\left\langle a^{\dagger}a\left(a+a^{\dagger}\right)+\left(a+a^{\dagger}\right)a^{\dagger}a\right\rangle}{2}-0, \\
            &= 0.
    \end{aligned}
\end{equation}

Thus we have
\begin{equation}
    \mathrm{var}\left(\mathcal{V}|_{\theta=0}\right) = \frac{t^4\eta^2}{4}(2n+1).
\end{equation}
This tells us that if we begin in a Fock state and drive the cavity at resonance, our ability to detect fluctuations in the cavity frequency grows quadratically with driving amplitude, linearly with photon number, and to the fourth power of time.

\subsubsection{Coherent state}
Let us now suppose the cavity is initially in a coherent state $|\alpha\rangle$. We have expectation values
\begin{equation}
    \begin{aligned}
        \langle 1\rangle &= 1, \\
        \langle a+a^{\dagger}\rangle &= \alpha+\alpha^*, \\
        \langle a^{\dagger}a\rangle &= |\alpha|^2.
    \end{aligned}
\end{equation}
The cross terms are:
\begin{equation}
    \begin{aligned}
        \langle 1^2\rangle &= 1. \\
        \langle 1(a+a^{\dagger})\rangle &= \alpha+\alpha^*. \\
        \langle 1(a^{\dagger}a)\rangle &= |\alpha|^2. \\
        \langle(a+a^{\dagger})^2\rangle &= \langle a^2+(a^{\dagger})^2+2a^{\dagger}a+1\rangle, \\
            &= (\alpha+\alpha^*)^2+1. \\
        \langle (a+a^{\dagger})a^{\dagger}a\rangle &= \langle a^{\dagger}a^2+(a^{\dagger})^2a+a\rangle, \\
            &= (\alpha+\alpha^*)|\alpha|^2+\alpha. \\
        \langle a^{\dagger}a(a+a^{\dagger})\rangle &= (\alpha+\alpha^*)|\alpha|^2+\alpha^*, \\
        \langle (a^{\dagger}a)^2\rangle &= \langle (a^{\dagger})^2a^2+a^{\dagger}a\rangle, \\
            &= |\alpha|^2\left(1+|\alpha|^2\right).
    \end{aligned}
\end{equation}

We thus have variances
\begin{equation}
    \begin{aligned}
        \mathrm{var}(1) &= \langle 1^2\rangle-\langle 1\rangle^2=0, \\
        \mathrm{var}(a+a^{\dagger}) &= \left((\alpha+\alpha^*)^2+1\right)-(\alpha+\alpha^*)^2=1, \\
        \mathrm{var}(a^{\dagger}a) &= |\alpha|^2\left(1+|\alpha|^2\right)-|\alpha|^4=|\alpha|^2,
    \end{aligned}
\end{equation}
and covariances
\begin{equation}
    \begin{aligned}
        \mathrm{covar}(1,X) &= 0\;\text{for all operators $X$,} \\
        \mathrm{covar}(a+a^{\dagger},a^{\dagger}a) &= (\alpha+\alpha^*)\left(|\alpha|^2+1/2\right)-|\alpha|^2(\alpha+\alpha^*), \\
            &= (\alpha+\alpha^*)/2.
    \end{aligned}
\end{equation}

The quantum Fisher information is then
\begin{equation}
    \begin{aligned}
        \mathrm{var}\left(\mathcal{V}|_{\theta=0}\right) &= \left(\frac{t^2\eta}{2}\right)^2+t^2|\alpha|^2-2\frac{t^3\eta}{2}\frac{\alpha+\alpha^*}{2}, \\
            &= \frac{\eta^2}{4}t^4-\frac{\eta(\alpha+\alpha^*)}{2}t^3+|\alpha|^2t^2.
            %&= \frac{t^2}{4}\left(4|\alpha|^2-(\alpha+\alpha^*)t+\eta^2t^2\right).
    \end{aligned}
\end{equation}
For small times this is approximately $|\alpha|^2t^2$, and the quantum Fisher information depends primarily on the coherent state amplitude and measurement time. In the long time limit however the initial coherent state amplitude ceases to be relevant, and the Fisher information is dominated by the square of the driving and the fourth power of time.

%Secondly for small $t$:
%\begin{equation}
%    i\left(\partial_{\theta}e^{-iHt}\right)e^{iHt}\right\rvert=\frac{i}{2}\eta t^2\left(a+a^{\dagger}\right)-it a^{\dagger}a+\mathcal{O}(t^3).
%\end{equation}

\subsection{Fisher information for time-dependent Hamiltonian}
Let us now suppose that the driving is varied sinusoidally with frequency $\omega_d$:
\begin{equation}
    h_1=0,h_2=-i\eta\sin(\omega_dt),\;h_3=i\eta\sin(\omega_dt),h_4=\theta.
\end{equation}
The quantum Fisher information vector evolves as:
\begin{equation}
    \dot{v}(t)=\partial_{\theta}h_j(\theta,t)+i\Gamma_{kl}^jv_k(t)h_l(\theta,t).
\end{equation}
Mathematica can solve this exactly for arbitrary $\theta$. However the resulting expressions are quite complicated, so we will just show $\theta=0$:
\begin{equation}
    \begin{aligned}
        v_1(t) &= \frac{\eta^2}{4\omega_d^3}\left[2\omega_dt\left(2+\cos(2\omega_dt)\right)-3\sin(2\omega_dt)\right], \\
        v_2(t) &=  \frac{\eta}{\omega_d^2}\left[\omega_dt\cos(\omega_dt)-\sin(\omega_dt)\right], \\
        v_3(t) &=  \frac{\eta}{\omega_d^2}\left[\omega_dt\cos(\omega_dt)-\sin(\omega_dt)\right], \\
        v_4(t) &=  t.
    \end{aligned}
\end{equation}

\subsection{The Wei-Norman method applied to a single cavity}

% The Wei-Norman expansion works quite well for this system. We parameterise the time-evolution operator as:
% \begin{equation}
%     U(t)=e^{u_1(t)}e^{u_2(t)a}e^{u_3(t)a^{\dagger}}e^{u_4(t)a^{\dagger}a}.
% \end{equation}
% The equations of motion are quite simple
% \begin{equation}
%     \begin{aligned}
%         \dot{u}_1(t) &= -\eta u_2(t), \\
%         \dot{u}_2(t) &= -\eta+i\Delta u_2(t), \\
%         \dot{u}_3(t) &= \eta-i\Delta u_3(t), \\
%         \dot{u}_4(t) &= -i\Delta.
%     \end{aligned}
% \end{equation}
% When the detuning and driving are constant, these have solution
% \begin{equation}
%     \begin{aligned}
%         u_1(t) &= \frac{\eta^2}{\Delta^2}\left(1-e^{it\Delta}+it\Delta\right), \\
%         u_2(t) &= -i\frac{\eta}{\Delta}\left(1-e^{it\Delta}\right), \\
%         u_3(t) &= -i\frac{\eta}{\Delta}\left(1-e^{-it\Delta}\right), \\
%         u_4(t) &= -it\Delta.
%     \end{aligned}
% \end{equation}

We will now solve the system using the Wei-Norman expansion. We begin by writing the time-evolution operator as a product of exponentials of the Lie algebra basis:
\begin{equation}
    U(t)=e^{-iF_1(t)a^\dagger a}e^{-iF_2(t)a^\dagger}e^{-iF_3(t)a}e^{-iF_4(t)},
\end{equation}
where the condition $U(0)=I$ gives initial condition $F_j(0)=0$. We now differentiate this as:
\begin{equation}
    \begin{aligned}
        \frac{dU}{dt}U^{-1}=&-i\dot{F_1}(t)a^\dagger a-i\dot{F_2}(t)U_1a^\dagger U_1^{-1}-\\
        &-i\dot{F_3}(t)U_1U_2aU_2^{-1}U_1^{-1}-i\dot{F_4}(t)=-iH.
     \end{aligned}
\end{equation}

Following the same procedure described in \cref{supp:SpinSystem}D, we need to compute the sandwich terms using the Baker-Campbell-Hausdorff formula:
\begin{equation}
    \begin{aligned}
        U_1a^\dag U_1^{-1}&=a^\dag+[-iF_1(t)a^\dag a,a^\dag]+\\
        &+\frac{1}{2}[-iF_1(t)a^\dag a,[-iF_1(t)a^\dag a,a^\dag]+...\\
        &=a^\dag e^{-iF_1(t)},\\
        U_1U_2aU_2^{-1}U_1^{-1}&=U_1(a+iF_2(t))U_1^{-1}=iF_2(t)+ae^{iF_1(t)}.
    \end{aligned}
\end{equation}
Using this we get the system of differential equations for the time-dependent functions $F_i(t)$:
\begin{equation}
    \begin{aligned}
        \dot{F_1}(t)&=\Delta ,\\
        \dot{F_2}(t)&=i\eta e^{iF_1(t)},\\
        \dot{F_3}(t)&=-i\eta e^{-iF_1(t)},\\
        \dot{F_4}(t)&=-F_2(t)\dot{F_3}(t).
    \end{aligned}
\end{equation}
These have solution:
\begin{equation}
    \begin{aligned}
        F_1(t) &=\Delta t,\\
         F_2(t)&=\eta/\Delta(e^{i\Delta t}-1),\\
         F_3(t)&=\eta/\Delta(e^{-i\Delta t}-1),\\
         F_4(t) &=-\eta^2/\Delta(t-ie^{-i\Delta t}/\Delta)+\eta^2/\Delta^2.
    \end{aligned}
\end{equation}
To reconstruct the time-evolution operator, we substitute the functions $F_i(t)$ back into the $U(t)$. For an easier interpretation of the evolution operator acting on the coherent states, we can combine operators $a,a^\dag$ to form a displacement operator:
\begin{equation}
    e^{-iF_2(t)a^\dag}e^{-iF_3(t)a}=e^{\xi a^\dag-\xi a+\eta^2/\Delta^2(1-\cos(\Delta t))},
\end{equation}
where we have defined $\xi=-i\eta/\Delta(e^{i\Delta t}-1)$. The resulting time evolution operator then takes the form:
\begin{equation}\label{eq:WNevolution}
    U(t)=e^{-i\Delta t a^\dagger a}e^{\xi a^\dagger -\xi^* a}e^{i\eta^2t/\Delta-i\eta^2/\Delta^2\sin{\Delta t}}.
\end{equation}

Let us calculate the QFI for estimating the cavity detuning $\Delta$ for a coherent initial state $|\alpha\rangle$. Then using our time-evolution operator the final state is:
\begin{equation}                
    |\psi_f\rangle=e^{i\eta^2t/\Delta+\eta^2/\Delta^2(e^{i\Delta t}-1)}|\alpha e^{-i\Delta t}-i\eta/\Delta(1-e^{-i\Delta t})\rangle.
\end{equation}
The Quantum Fisher Information of a parameter $\theta$ encoded in a pure state $|\psi\rangle$ is given by \cite{paris_quantum_2011}: 
\begin{equation}
    F_{\theta}=4(\langle\partial_{\theta}\psi|\partial_{\theta}\psi\rangle-|\langle\partial_{\theta}\psi|\psi\rangle|^2).
\end{equation}
We thus must compute the derivative of $|\psi_f\rangle$ with respect to $\Delta$. This step involves a significant amount of operator re-arranging, and yields:
\begin{equation}
    \begin{aligned}
        |\partial_{\Delta}\psi_f\rangle &= \partial_{\Delta} U|\alpha\rangle, \\
        &= e^{-i\Delta t a^\dagger a}e^{\xi a^\dagger -\xi^* a}e^{i\eta^2t/\Delta-i\eta^2/\Delta^2\sin{\Delta t}}\times\\
        &[-it(a^\dag+\xi^*)(a+\xi)+\partial\xi a^\dag-\partial\xi^* a+\partial \Gamma(\Delta)]|\alpha\rangle.
    \end{aligned}
\end{equation}

Next, when substituting this expression in the QFI formula we require the mean values for the operators $a,a^\dag,a^\dag a, (a^\dag a)^2,(a^\dag)^2,a^2$. These were previously computed in \cref{app:SingleCavity}C for our linear parameterisation. The Quantum Fisher information for coherent states is then
\begin{equation}
    \begin{aligned}
    \frac{F_Q}{4} &= t^2(|\alpha|^2 +|\xi|^2 +\alpha^*\xi+\alpha\xi^*) +|\partial_{\Delta}\xi|^2 +\\
    &+it(\alpha^*\partial_{\Delta}\xi-\alpha\partial_{\Delta}\xi^* +\xi^*\partial_{\Delta}\xi -\xi\partial_{\Delta}\xi^*).
    \end{aligned}
\end{equation}
We can evaluate:
\begin{equation}
    \begin{aligned}
    |\partial_{\Delta}\xi|^2 &=\frac{2\eta^2}{\Delta^4}(1-\cos(\Delta t))-\frac{2\eta^2t}{\Delta^3}\sin(\Delta t)+\frac{\eta^2t^2}{\Delta^2}, \\
    |\xi|^2 &= 2t^2\eta^2/\Delta^2(1-\cos(\Delta t)).
    \end{aligned}
\end{equation}
Taking the limit $\Delta\rightarrow0$ gives:
\begin{equation}\label{eq:WNQFIcoherent}
    \frac{F_Q}{4}\rvert_{\Delta=0}=t^2|\alpha|^2+\eta^2t^4/4-(\alpha+\alpha^*)\eta t^3/2
\end{equation}

By following the same process as above, we can calculate the QFI for Fock states $|n\rangle$:
\begin{equation}
    \frac{F_Q}{4}=\eta^2/\Delta^4(\Delta^2t^2+2(1-\cos(\Delta t))-2\Delta t(\sin(\Delta t)))(2n+1).
\end{equation}
Taking the limit $\Delta\rightarrow0$ gives:
\begin{equation}\label{eq:WNQFIfock}
    \frac{F_Q}{4}\rvert_{\Delta=0}=(2n+1)\eta^2t^4/4
\end{equation}

Equations \cref{eq:WNQFIcoherent,eq:WNQFIfock} yield the same result as our linear parameterisation in \cref{app:SingleCavity}C.

\section{Nonlinear optomechanics}\label{supp:NonlinearOptomech}
The Wei Norman expansion was used in \cite{schneiter_optimal_2020,qvarfort_constraining_2022,qvarfort_solving_2022} to solve the nonlinear optomechanical Hamiltonian, and study the Fisher information of various parameters. In this section, we will show how this same problem can be approached using linear Lie algebra parameterisations.

The optomechanical Hamiltonian is
\begin{equation}
    H_{OM}=\hbar\omega a^{\dagger}a+\frac{p^2}{2m}+\frac{1}{2}m\Omega(t)^2x^2-\hbar G(t) a^{\dagger}a x+mg(t)\,x.
\end{equation}
This is equivalent to that from \cite{schneiter_optimal_2020}, using the position and momentum rather than phonon basis for the mechanical mode as a matter of preference. The optical field has annihilation operator $a$ and frequency $\omega$. The mechanical mode has position $x$ and momentum $p$, and mass $m$. The mechanical frequency is $\Omega(t)$, which can be time-dependent. The optomechanical coupling is $G(t)$. We allow for gravitational coupling through $g(t)$. Note that this Hamiltonian omits a driving term, which would appear as $i\eta(a^{\dagger}-a)$. If the operators $a,a^{\dagger}$ were included as generators, the Lie algebra generated by the Hamiltonian would be infinite. Tackling infinite-dimensional Lie algebras is a topic of ongoing research.

%The optomechanical Hamiltonian was defined as \cref{sc}\cite[\S II]{schneiter_optimal_2020}
%\begin{equation}
%    H_{OM}=\hbar\omega_c a^{\dagger}a+\hbar\omega_m b^{\dagger}b-\hbar\mathcal{G}(t)a^{\dagger}a(b^{\dagger}+b).
%\end{equation}
%Here the optical field has annihilation operator $a$, and frequency $\omega_c$ is the optical frequency. The mechanical mode has phonon annihilation operator $b$, and frequency $\omega_m$ the mechanical frequency. The optomechanical coupling is $\mathcal{G}(t)$, is allowed to be time-dependent. Re-scaling this
%\begin{equation}
%    \begin{aligned}
%        H/(\hbar\omega_m) &= \tilde{H}_{OM}+\tilde{\mathcal{D}}_1(\tau)(b^{\dagger}+b)+\tilde{\mathcal{D}}_2(\tau)(b^{\dagger}+b)^2, \\
%        \tilde{H}_{OM} &= \Omega a^{\dagger}a+b^{\dagger}b-\tilde{\mathcal{G}}(\tau)a^{\dagger}a(b^{\dagger}+b).
%    \end{aligned}
%\end{equation}

\subsection{Lie algebra}
The Hamiltonian contains the operators
\begin{equation}
    x,x^2,p^2,n,nx,
\end{equation}
where $n=a^{\dagger}a$ is the number operator. By taking commutators of these elements, we need the following elements to complete the Lie algebra:
\begin{equation}
    1,p,xp,np,n^2.
\end{equation}

\begin{widetext}
This gives us a ten-dimensional Lie algebra, with a commutator table:
\begin{center}
    \begin{tabular}{c|cccccccccc}
     $[X,Y]/\hbar$   & $1$ & $n$ & $n^2$ & $x$ & $p$ & $p^2$ & $x^2$ & $xp$ & $nx$ & $np$ \\
    \hline
    $1$  \\
    $n$ \\
    $n^2$ \\
    $x$   & & & & &$i $&$2i p$& &$i  x$& &$i  n$\\ 
    $p$   & & & &$-i $& & &$-2i x$&$-i p$&$-i n$&\\ 
    $p^2$ & & & &$-2i p$& & &$-2(\hbar +2ixp)$&$-2i p^2$&$-2i np$& \\ 
    $x^2$ & & & & &$2i x$&$2(\hbar +2 ixp)$& &$2i x^2$& &$2i nx$\\ 
    $xp$  & & & &$-i x$&$i p$&$2i p^2$&$-2i x^2$& &$-i nx$&$i np$\\ 
    $nx$  & & & & &$i n$&$2i np$& &$i nx$& &$i n^2$\\ 
    $np$  & & & &$-i n$& & &$-2i nx$&$-i np$&$-i n^2$& \\ 
    \end{tabular}
\end{center}
Note that we have divided each commutator by $\hbar$. From this table we could compute the adjoint representation, then solve the dynamics in the Heisenberg picture. However since the Lie algebra is ten-dimensional, the linear Heisenberg equations of motion would be 100 coupled differential equations. Mathematica can solve these equations in the case where $\Omega,G,g$ are time-independent. However, we can greatly simplify the analysis by working on a lower-dimensional Lie algebra ideal, as discussed in \cref{sec:Ideals} of the main text.

We are particularly interested in the dynamics of the operators $\{x,p,n\}$. These tell us about the position and momentum of the oscillator, and the number of photons in the cavity. Moreover, by taking products of these we can compute all the other operators in the Lie algebra. By looking at the first few columns of the commutator table, we can see that the set $\mathcal{I}=\{1,x,p,n\}$ forms an ideal, in other words $[X_j,I_k]\in\mathcal{I}$ for all $X_j\in\mathcal{L},I_k\in\mathcal{I}$.

%\begin{widetext}
Let us construct the adjoint representation restricted to the ideal. Then for any $X_j$ in the original Lie algebra, $\mathrm{ad}_{X_j}$ will be only a $4\times 4$ matrix. The $kl$th element of this will be the $X_k$ component of $\mathrm{ad}_{X_j}X_l$. We can read these off the commutator table:
\begin{equation}
    \begin{gathered}
        \mathrm{ad}_I=
        \left(
        \begin{array}{cccc}
         0 & 0 & 0 & 0 \\
         0 & 0 & 0 & 0 \\
         0 & 0 & 0 & 0 \\
         0 & 0 & 0 & 0 \\
        \end{array}
        \right),\;
        \mathrm{ad}_n=
        \left(
        \begin{array}{cccc}
         0 & 0 & 0 & 0 \\
         0 & 0 & 0 & 0 \\
         0 & 0 & 0 & 0 \\
         0 & 0 & 0 & 0 \\
        \end{array}
        \right),\;
        \mathrm{ad}_{n^2}=
        \left(
        \begin{array}{cccc}
         0 & 0 & 0 & 0 \\
         0 & 0 & 0 & 0 \\
         0 & 0 & 0 & 0 \\
         0 & 0 & 0 & 0 \\
        \end{array}
        \right),\\
        \mathrm{ad}_{x}=
        \left(
        \begin{array}{cccc}
         0 & 0 & 0 & i \\
         0 & 0 & 0 & 0 \\
         0 & 0 & 0 & 0 \\
         0 & 0 & 0 & 0 \\
        \end{array}
        \right),\;
        \mathrm{ad}_{p}=
        \left(
        \begin{array}{cccc}
         0 & 0 & -i & 0 \\
         0 & 0 & 0 & 0 \\
         0 & 0 & 0 & 0 \\
         0 & 0 & 0 & 0 \\
        \end{array}
        \right),\; \\
        \mathrm{ad}_{x^2}=
        \left(
        \begin{array}{cccc}
         0 & 0 & 0 & 0 \\
         0 & 0 & 0 & 0 \\
         0 & 0 & 0 & 2i \\
         0 & 0 & 0 & 0 \\
        \end{array}
        \right),\;
        \mathrm{ad}_{p^2}=
        \left(
        \begin{array}{cccc}
         0 & 0 & 0 & 0 \\ 
         0 & 0 & 0 & 0 \\
         0 & 0 & 0 & 0 \\
         0 & 0 & -2i & 0 \\
        \end{array}
        \right),\;
        \mathrm{ad}_{xp}=
        \left(
        \begin{array}{cccc}
         0 & 0 & 0 & 0 \\
         0 & 0 & 0 & 0 \\
         0 & 0 & -i & 0 \\
         0 & 0 & 0 & i \\
        \end{array}
        \right),\; \\
        \mathrm{ad}_{nx}=
        \left(
        \begin{array}{cccc}
         0 & 0 & 0 & 0 \\
         0 & 0 & 0 & i \\
         0 & 0 & 0 & 0 \\
         0 & 0 & 0 & 0 \\
        \end{array}
        \right),\;
        \mathrm{ad}_{np}=
        \left(
        \begin{array}{cccc}
         0 & 0 & 0 & 0 \\
         0 & 0 & -i & 0 \\
         0 & 0 & 0 & 0 \\
         0 & 0 & 0 & 0 \\
        \end{array}
        \right).
    \end{gathered}
\end{equation}
%\end{widetext}

%\begin{widetext}
\subsection{Dynamics using linear parameterisation}
Let us first solve this in the time-independent case, where $g(t)=g,\Omega(t)=\Omega, G(t)=G$. On the full algebra, the Lie algebra Hamiltonian matrix \cref{eq:LieHeisenbergEOM} is
\begin{equation}\label{eq:OptomechHMatrix}
    \mathbf{H} =
    \left(
\begin{array}{cccccccccc}
 0 & 0 & 0 & 0 & 0 & 0 & 0 & 0 & 0 & 0 \\
 0 & 0 & 0 & 0 & 0 & 0 & 0 & 0 & 0 & 0 \\
 0 & 0 & 0 & 0 & 0 & 0 & 0 & 0 & 0 & 0 \\
 0 & 0 & 0 & 0 & -\frac{i \hbar }{m} & 0 & 0 & 0 & 0 & 0 \\
 i g m \hbar  & -i G \hbar ^2 & 0 & i m \Omega ^2 \hbar  & 0 & 0 & 0 & 0 & 0 & 0 \\
 m \Omega ^2 \hbar ^2 & 0 & 0 & 0 & 2 i g m \hbar  & 0 & 0 & 2 i m \Omega ^2 \hbar  & 0 & -2 i G \hbar ^2 \\
 -\frac{\hbar ^2}{m} & 0 & 0 & 0 & 0 & 0 & 0 & -\frac{2 i \hbar }{m} & 0 & 0 \\
 0 & 0 & 0 & i g m \hbar  & 0 & -\frac{i \hbar }{m} & i m \Omega ^2 \hbar  & 0 & -i G \hbar ^2 & 0 \\
 0 & 0 & 0 & 0 & 0 & 0 & 0 & 0 & 0 & -\frac{i \hbar }{m} \\
 0 & i g m \hbar  & -i G \hbar ^2 & 0 & 0 & 0 & 0 & 0 & i m \Omega ^2 \hbar  & 0 \\
\end{array}
\right)
\end{equation}

Using Mathematica, we can solve the Lie-algebra equation of motion \cref{eq:LieHeisenbergEOM} analytically and obtain:
\begin{equation}\label{eq:NonlinearOptomechSolution}
    \begin{aligned}
        n(t)    &= n, \\
        x(t)    &= -\frac{2g}{\Omega^2}\sin^2(\Omega t/2)+\frac{2\hbar G}{m\Omega^2}\sin^2(\Omega t/2)n+\cos(\Omega t)x+\frac{1}{m\Omega}\sin(\Omega t)p, \\
        p(t)    &= -\frac{mg}{\Omega}\sin(\Omega t)+\frac{\hbar G}{\Omega}\sin(\Omega t)n-m\Omega\sin(\Omega t)x+\cos(\Omega t)p.
    \end{aligned}
\end{equation}
All other operators in the Lie algebra can be constructed from products or sums of these. This therefore gives a globally valid solution for the full nonlinear quantum dynamics.
%\end{widetext}

In fact, we need not use the full algebra. The photon number $n$ is preserved by these dynamics. We thus need only solve for $x(t)$, since we can derive $p(t)$ by differentiating this. Then by taking products we can generate the entire Lie algebra. We can thus focus only on the set of operators generated by repeated commutators of $H$ with $X$.

Suppose we wish to consider the evolution of $x$. 
\begin{enumerate}
    \item Taking the commutator $[\mathcal{H},x]$ produces the operators $\{x,p\}$.
    \item Taking another commutator $[\mathcal{H},\{x,p\}]$ produces the operators $\{1,n\}$.
    \item Taking successive commutators does not generate any new operators.
\end{enumerate}
Therefore evolution of the $x$ operator is restricted to the ideal:
\begin{equation}
    \mathcal{X}=\{1,n,x,p\}.
\end{equation}
The Hamiltonian matrix in the ideal is:
\begin{equation}\label{eq:IdealHamiltonian}
    \mathbf{H}=
    i\hbar\left(
\begin{array}{cccc}
 0 & 0 & 0 & 0 \\
 0 & 0 & 0 & 0 \\
 0 & 0 & 0 & -1/m \\
   mg(t) & - G(t)  ^2 &  m   \Omega(t)^2 & 0 \\
\end{array}
\right)
\end{equation}
This reduces the number of differential equations from $10^2=100$ to $4^2=16$. Solving the equations of motion \cref{eq:LieHeisenbergEOM} using this matrix gives us the same solution \cref{eq:NonlinearOptomechSolution}.

%Using \cref{eq:LieHeisenbergEOM} we derive the following four coupled systems:
%\begin{equation}
%    \begin{aligned}
%        u_{3,1}' &= \frac{u_{4,1}}{m},\;  u_{4,1}' = -m(\Omega+\mathcal{D}_2)^2u_{3,1}-(mg+\mathcal{D}_1), \\
%        u_{3,2}' &= \frac{u_{4,2}}{m},\;  u_{4,2}' = -m(\Omega+\mathcal{D}_2)^2u_{3,2}+\hbar G, \\
%        u_{3,3}' &= \frac{u_{4,3}}{m},\; u_{4,3}' = -m(\Omega+\mathcal{D}_2)^2u_{3,3}, \\
%        u_{3,4}' &= \frac{u_{4,4}}{m},\; u_{4,4}' = -m(\Omega+\mathcal{D}_2)^2u_{3,4}.
%    \end{aligned}
%\end{equation}

We will now apply the linear parameterisations to compute the dynamics in the three cases of time-varying Hamiltonian studied in \cite{schneiter_optimal_2020}. We reiterate that it is not strictly necessary to find the dynamics if all we require is the quantum Fisher information. However, explicit Heisenberg-picture dynamics can be more illuminating than the product of exponentials resulting from the Wei-Norman expansion.

%\begin{widetext}
\subsubsection{Time-dependent gravitational signal}
Suppose the gravitational field varies as
\begin{equation}
    g(t)=g_0\left(1+\theta\cos(\omega_gt)\right).
\end{equation}
Substituting this into \cref{eq:IdealHamiltonian} and solving the Fernández equations of motion \cref{eq:LieHeisenbergEOM} gives
%\begin{widetext}
\begin{equation}
    \begin{aligned}
        x(t) &= \frac{-2g_0}{\Omega^2}\left(\sin^2(\Omega t/2)+\theta\frac{\Omega^2}{2}\frac{\cos(\omega_g t)-\cos(\Omega t)}{\Omega^2-\omega_g^2}\right)+\frac{2\hbar G}{m\Omega^2}\sin^2(\Omega t/2)n+\cos(\Omega t)x+\frac{1}{m\Omega}\sin(\Omega t)p, \\
        p(t) &= -\frac{mg_0}{\Omega}\left(\sin(\Omega t)+\theta\Omega\frac{\Omega\sin(\Omega t)-\omega_g\sin(\omega_g t)}{\Omega^2-\omega_g^2}\right)+\frac{\hbar G}{\Omega}\sin(\Omega t)n-m\Omega\sin(\Omega t)x+\cos(\Omega t)p.
    \end{aligned}
\end{equation}

The number operator is preserved by these dynamics: $n(t)=n$. Taking products of these operators is sufficient to reconstruct the time-evolution of the entire algebra. Note that terms on the right-hand-side not proportional to an operator, i.e. the first terms after the equal sign, are proportional to the identity.

This is a fully quantum solution to the dynamics. We can see that the effect of the gravitational signal is simply to add a harmonic term to the position.

\subsubsection{Time-dependent optomechanical coupling}
Now suppose that it is the optomechanical coupling is time-dependent:
\begin{equation}\label{eq:VaryingOMCoupling}
    G(t)=G_0\left(1+\theta\cos(\omega_G t)\right).
\end{equation}
Substituting this into \cref{eq:IdealHamiltonian}, the equations of motion \cref{eq:LieHeisenbergEOM} can be immediately solved using Mathematica:
\begin{equation}
    \begin{aligned}
        x(t) &= -\frac{2g}{\Omega^2}\sin^2(\Omega t/2)+ \frac{2\hbar G_0}{m\Omega^2}\left(\sin^2(\Omega t/2)+\frac{\theta\Omega^2}{2}\frac{\cos(\omega_G t)-\cos(\Omega t)}{\Omega^2-\omega_G^2}\right)n+\cos(\Omega t)x+\frac{1}{m\Omega}\sin(\Omega t)p, \\
        p(t) &= -\frac{mg}{\Omega}\sin(\Omega t)+\frac{\hbar G_0}{\Omega}\left(\sin(\Omega t)+\theta\Omega\frac{\Omega\sin(\Omega t)-\omega_G\sin(\omega_Gt)}{\Omega^2-\omega_G^2}\right)n-m\Omega\sin(\Omega t)x+\cos(\Omega t)p.
    \end{aligned}
\end{equation}

We can see that the effect is simply to add a harmonic term to the photon number $n$.
\end{widetext}
%\end{widetext}

\subsubsection{Time-dependent mechanical frequency}
Finally let us consider varying the mechanical frequency:
\begin{equation}
    \Omega(t)^2=m\left(\Omega_0^2+\theta\cos(\Omega_mt)\right).
\end{equation}
This is more challenging than the previous two examples. The identity and number operator are conserved, which gives us:
\begin{equation}
    \begin{gathered}
        u_{11}(t)=1,\;u_{12}(t)=0,\;u_{13}(t)=0,\;u_{14}(t)=0, \\
        u_{21}(t)=0,\;u_{22}(t)=1,\;u_{23}(t)=0,\;u_{24}(t)=0.
    \end{gathered}
\end{equation}
The operators for $x$ satisfy
\begin{equation}
    \dot{u}_{31} = \frac{u_{41}}{m},\; \dot{u}_{32} = \frac{u_{42}}{m},\; \dot{u}_{33} = \frac{u_{43}}{m},\; \dot{u}_{34} = \frac{u_{44}}{m},
\end{equation}
in analogy to $\dot{x}=p/m$:
For $p$ we have a system analogous to $\dot{p}=\hbar G|\alpha|^2-m\Omega^2x-mg$:
\begin{equation}
    \dot{u}_{4j}(t) =-mgu_{1j}(t)+ \hbar G u_{2j}(t)-m\left(\Omega_0^2+\theta\cos(\omega_mt)\right)u_{3j}(t).
\end{equation}
We have $u_{1j}(t)=\delta_{j1}$ and $u_{2j}(t)=\delta_{j2}$. Thus we get four systems of two coupled equations:
\begin{align}
    \dot{u}_{31} &= u_{41}/m,   &&\dot{u}_{41} = -m\left(\Omega_0^2+\theta\cos(\omega_mt)\right)u_{31}-mg, \label{eq:Mathieu1} \\
    \dot{u}_{32} &= u_{42}/m,   &&\dot{u}_{42} = -m\left(\Omega_0^2+\theta\cos(\omega_mt)\right)u_{32}+\hbar G, \label{eq:Mathieu2} \\
    \dot{u}_{33} &= u_{43}/m,   &&\dot{u}_{43} = -m\left(\Omega_0^2+\theta\cos(\omega_mt)\right)u_{33}, \label{eq:Mathieu3} \\
    \dot{u}_{34} &= u_{44}/m,   &&\dot{u}_{44} = -m\left(\Omega_0^2+\theta\cos(\omega_mt)\right)u_{34}, \label{eq:Mathieu4}
\end{align}
with initial conditions $u_{3j}(0)=\delta_{j3}$, and $u_{4j}(0)=\delta_{j4}$. We can recognize these systems as Mathieu's equations. Mathieu's equation also appeared in \cite{schneiter_optimal_2020} using the Wei-Norman parameterisation. It is interesting that the same system turns up in both approaches. There is likely fruitful research to be done on the precise relationship between the Wei-Norman and linear Heisenberg parameterisations.

The equations \cref{eq:Mathieu3} and \cref{eq:Mathieu4} can be solved by Mathematica. 
However, Mathematica is unable to deal with \cref{eq:Mathieu1,eq:Mathieu2}. We must therefore roll up our sleeves, and solve these systems ourselves using the theory of Mathieu equations.
%. The differential equations are identical, differing only in their initial conditions. If we simply stick these into Mathematica, we can arrive at a solution for the first two:
%\begin{align}
%    u_{34}(t) &= \frac{2 S\left(\frac{4 \Omega_0^2}{\omega_m^2},-\frac{2 \theta }{\omega_m^2},\frac{\omega_mt}{2}\right)}{m \omega_m S'\left(\frac{4 \Omega_0^2}{\omega_m^2},-\frac{2 \theta }{\omega_m^2},0\right)},\; &&u_{44}(t)=\frac{S'\left(\frac{4 \Omega_0^2}{\omega_m^2},-\frac{2 \theta }{\omega_m^2},\frac{\omega_mt}{2}\right)}{S'\left(\frac{4 \Omega_0^2}{\omega_m^2},-\frac{2 \theta }{\omega_m^2},0\right)}, \\
%    u_{33}(t) &= \frac{C\left(\frac{4 \Omega_0^2}{\omega_m^2},-\frac{2 \theta }{\omega_m^2},\frac{\omega_mt}{2}\right)}{C\left(\frac{4 \Omega_0^2}{\omega_m^2},-\frac{2 \theta }{\omega_m^2},0\right)},\; &&u_{43}(t)=\frac{m \omega_m C'\left(\frac{4 \Omega_0^2}{\omega_m^2},-\frac{2 \theta }{\omega_m^2},\frac{\omega_mt}{2}\right)}{2 C\left(\frac{4 \Omega_0^2}{\omega_m^2},-\frac{2 \theta }{\omega_m^2},0\right)},
%\end{align}
%where $S$ and $C$ are Mathieu's functions. 

Mathieu's equation is
\begin{equation}\label{eq:MathieusEquation}
    \frac{\mathrm{d}^2u}{\mathrm{d}t^2}=-\left(a-2q\cos(2t)\right)u,
\end{equation}
for constants $a$ and $q$. This has two linearly independent solutions, $S(a,q,t)$ and $C(a,q,t)$, which we can imagine as generalized sines and cosines. In analogy to sine we have $S(a,q,0)=0$, and $S$ is an odd function. In analogy to cosine we have $C(a,q,0)=1$ and $C$ is an even function.

All of the equations \cref{eq:Mathieu1,eq:Mathieu2,eq:Mathieu3,eq:Mathieu4} are of the form:
\begin{equation}\label{eq:MathieuGeneralForm}
    \dot{u}_{3j}=\frac{u_{4j}}{m},\;\dot{u}_{4j}=-m\left(\Omega_0^2+\theta\cos(\omega_mt)\right)u_{3j}+F_j,
\end{equation}
for $F_j=(-mg,\hbar G,0,0)$. Let us massage these into the form of \cref{eq:MathieusEquation}. Differentiating the second equation, and using $\dot{F}_j=0$, we find:
\begin{equation}
    \ddot{u}_{4j}=-\left(\Omega_0^2+\theta\cos(\omega_mt)\right)u_{4j}.
\end{equation}
This is almost of the form \cref{eq:MathieusEquation}, except for the argument of $\omega_mt$ in the cosine. To deal with this we rescale time as $\omega_mt=2\tau$, then $\partial_t=(\omega_m/2)\partial_{\tau}$. Letting primes denote derivatives with respect to $\tau$, we find:
\begin{equation}
    \begin{aligned}
        \frac{\omega_m^2}{4}u''_{4j} &= -\left(\Omega_0^2+\theta\cos(2\tau)\right)u_{4j}, \\
        u''_{4j} &= -\left[\left(\frac{4\Omega_0^2}{\omega_m^2}\right)-2\left(\frac{-2\theta}{\omega_m^2}\right)\cos(2\tau)\right]u_{4j}.
    \end{aligned}
\end{equation}
This is Mathieu's equation, with constants
\begin{equation}
    a=\frac{4\Omega_0^2}{\omega_m^2},\;q=\frac{-2\theta}{\omega_m^2}.
\end{equation}
The general solution can then be written as a linear combination of Mathieu functions:
\begin{equation}
    u_{4j}=\alpha\,C(a,q,\tau)+\beta S(a,q,\tau),
\end{equation}
or returning to our original timescale using $\tau=\omega_mt/2$:
\begin{equation}
    u_{4j}=\alpha\,C\left(\frac{4\Omega_0^2}{\omega_m^2},\frac{-2\theta}{\omega_m^2},\frac{\omega_mt}{2}\right)+\beta S\left(\frac{4\Omega_0^2}{\omega_m^2},\frac{-2\theta}{\omega_m^2},\frac{\omega_mt}{2}\right).
\end{equation}

The constants $\alpha,\beta$ are fixed by the initial conditions. We have
\begin{equation}
    \begin{aligned}
        u_{4j}(0)  &= \alpha\,C(a,q,0)+\beta S(a,q,0), \\
        \dot{u}_{4j}(0) &= \alpha\,\dot{C}(a,q,0)+\beta \dot{S}(a,q,0),
    \end{aligned}
\end{equation}
where the derivative is with respect to $t$. We must find $\alpha,\beta$ that can satisfy these. This can be written in matrix form:
\begin{equation}\label{eq:MathieuICMatrixRelation}
    \left(\begin{array}{c}u_{4j}(0) \\ \dot{u}_{4j}(0)\end{array}\right)=\left(\begin{array}{cc}C(a,q,0) & S(a,q,0) \\ \dot{C}(a,q,0) & \dot{S}(a,q,0)\end{array}\right)\left(\begin{array}{c}\alpha \\ \beta \end{array}\right).
\end{equation}
Inverting this then gives us:
\begin{equation}
    \alpha = \frac{u_{4j}(0)}{C(a,q,0)},\; \beta  = \frac{\dot{u}_{4j}(0)}{\dot{S}(a,q,0)}.
\end{equation}

We thus need two initial conditions on $u_{4j}$. We are given $u_{4j}(0)=\delta_{4j}$, where $\delta_{jk}$ is the Kronecker delta function. For the derivative we have:
\begin{equation}
    \dot{u}_{4j}(0)=-m\left(\Omega_0^2+\theta\right)u_{3j}(0)+F_j,
\end{equation}
where $u_{3j}(0)=\delta_{3j}$. Thus we find:
\begin{equation}
    \begin{aligned}
    \alpha &=\frac{\delta_{4j}}{C\left(\frac{4\Omega_0^2}{\omega_m^2},\frac{-2\theta}{\omega_m^2},0\right)}, \\
    \beta &= \frac{\dot{u}_{4j}(0)}{\dot{S}(a,q,0)}=\frac{-m\left(\Omega_0^2+\theta\right)\delta_{3j}+F_j}{\dot{S}\left(\frac{4\Omega_0^2}{\omega_m^2},\frac{-2\theta}{\omega_m^2},0\right)}.
    \end{aligned}
\end{equation}
Thus we find:

\begin{equation}
    \begin{aligned}
        u_{41} &= \frac{-mg}{\dot{S}(a,q,0)}S(a,q,\omega_mt/2) \\
        &= \frac{-mg}{\dot{S}\left(\frac{4\Omega_0^2}{\omega_m^2},\frac{-2\theta}{\omega_m^2},0\right)}S\left(\frac{4\Omega_0^2}{\omega_m^2},\frac{-2\theta}{\omega_m^2},\frac{\omega_mt}{2}\right), \\
        u_{42} &= \frac{\hbar G}{\dot{S}(a,q,0)}S(a,q,\omega_mt/2) \\
        &=\frac{\hbar G}{\dot{S}\left(\frac{4\Omega_0^2}{\omega_m^2},\frac{-2\theta}{\omega_m^2},0\right)}S\left(\frac{4\Omega_0^2}{\omega_m^2},\frac{-2\theta}{\omega_m^2},\frac{\omega_mt}{2}\right), \\
        u_{43} &=\frac{-m(\Omega_0^2+\theta)}{\dot{S}(a,q,0)}S(a,q,\omega_mt/2) \\
        &= \frac{-m\left(\Omega_0^2+\theta\right)}{\dot{S}\left(\frac{4\Omega_0^2}{\omega_m^2},\frac{-2\theta}{\omega_m^2},0\right)}S\left(\frac{4\Omega_0^2}{\omega_m^2},\frac{-2\theta}{\omega_m^2},\frac{\omega_mt}{2}\right), \\
        u_{44} &= \frac{1}{C(a,q,0)}C(a,q,\omega_mt/2) \\
        &= \frac{1}{C\left(\frac{4\Omega_0^2}{\omega_m^2},\frac{-2\theta}{\omega_m^2},0\right)}C\left(\frac{4\Omega_0^2}{\omega_m^2},\frac{-2\theta}{\omega_m^2},\frac{\omega_mt}{2}\right).
    \end{aligned}
\end{equation}

This gives us our solution for $p(t)$. Recall that $S$, in analogy to sine, is zero at $t=0$. Thus as expected initially only the momentum component of $u_{4j}$ is nonzero.

We can find $u_{3j}$, and hence $x(t)$, from \cref{eq:MathieuGeneralForm}:
\begin{equation}
    u_{3j}(t)=-\frac{\dot{u}_{4j}-F_j}{m\left(\Omega_0^2+\theta\cos(\omega_mt)\right)}.
\end{equation}

This gives us an exact solution for the dynamics. 

\subsection{Quantum Fisher information}
We will now compute the quantum Fisher information vector for the three aforementioned cases. The covariances of these with respect to an initial state can then be used to find the QFI, as described in Appendix A of the main text. In each of these cases the Hamiltonian will be time-dependent, so we will find $\mathcal{V}$ using \cref{eq:FisherFormulaTimeDependent}.

\subsubsection{Time-dependent gravitational signal}
Suppose the gravitational field contains a signal varying as
\begin{equation}
    g(t)=g_0\left(1+\theta\cos(\omega_gt)\right).
\end{equation}
We could compute all of the $\dot{v}_j$ and solve these as before. However in the Hamiltonian $g(t)$ is proportional to the $x$ operator, which generates a four-dimensional ideal $\{1,n,x,p\}$. Thus we need only consider the coefficients $v_1,v_2,v_4,v_5$, as all others must be zero. \cref{eq:FisherFormulaTimeDependent} gives
\begin{equation}
    \begin{aligned}
        \dot{v}_1(t) &= mg_0\left(1+\theta\cos(\omega_gt)\right)\,v_5(t) \\
            &\phantom{=}+i(\hbar/m)\left(v_6(t)-m^2\Omega^2\,v_7(t)\right), \\
        \dot{v}_2(t) &= -\hbar G\,v_5(t)+mg_0\left(1+\theta\cos(\omega_gt)\right)\,v_{10}(t), \\
        \dot{v}_4(t) &= (mg_0/\hbar)\cos(\omega_gt)+m\Omega^2\,v_5(t) \\
            &\phantom{=}+mg_0\left(1+\theta\cos(\omega_gt)\right)\,v_8(t), \\
        \dot{v}_5(t) &= -(1/m)\,v_4(t)+2mg_0\left(1+\theta\cos(\omega_gt)\right)\,v_7(t).
    \end{aligned}
\end{equation}
We must have $v_j(t)=0$ for $j$ not equal to $1,2,4,5$. Since $v_1(t)$ does not appear in the last three equations we need not solve for this, since as described in Appendix A of the main text it does not affect the quantum Fisher information. Thus we find:
\begin{equation}
    \begin{aligned}
        v_2(t) &= \frac{Gg_0}{\Omega\omega_g}\left(\frac{\Omega\sin(\omega_g t)-\omega_g\sin(\Omega t)}{\Omega^2-\omega_g^2}\right), \\
        v_4(t) &= \frac{mg_0}{\hbar}\frac{\Omega\sin(\Omega t)-\omega_g\sin(\omega_g t)}{\Omega^2-\omega_g^2}, \\
        v_5(t)&= \frac{g_0}{\hbar}\frac{\cos(\Omega t)-\cos(\omega_g t)}{\Omega^2-\omega_g^2}.
    \end{aligned}
\end{equation}
%For completeness we will include the solution for $v_1$:
%\begin{equation}
%    \begin{aligned}
%    v_1(t) &= \frac{m\theta}{4\hbar\omega_g\Omega^2-\omega_g^2)^2}\left[4\omega_g\left(\Omega\cos(\omega_gt)\sin(\Omega t)\right.\right. \\
%    &\left.\left.-\omega_g\sin(\omega_gt)\cos(\Omega t)\right)-(\Omega^2-\omega_g^2)\left(\sin(2\omega_gt)+2\omega_gt\right)\right].
%    \end{aligned}
%\end{equation}

\subsubsection{Time-dependent optomechanical coupling}
Now suppose the optomechanical coupling varies as:
\begin{equation}
    G(t)=\theta\left(1+\theta\cos(\omega_G t)\right).
\end{equation}
The coupling term $nx$ generates the ideal spanned by $\{n,n^2,nx,np\}$. We thus need only consider the coefficients $v_2,v_3,v_9,v_{10}$; the other coefficients will be identically zero. Then \cref{eq:FisherFormulaTimeDependent} gives
\begin{equation}
    \begin{aligned}
        \dot{v}_2(t) &= mg\,v_{10}(t), \\
        \dot{v}_3(t) &= -\hbar G_0\left(1+\theta\cos(\omega_Gt)\right)\,v_{10}(t), \\
        \dot{v}_9(t) &= -G_0\cos(\omega_Gt)+m\Omega^2\,v_{10}(t), \\
        \dot{v}_{10}(t) &= -v_9(t)/m.
    \end{aligned}
\end{equation}
%\begin{equation}
%    \begin{aligned}
%        \dot{v}_1(t) &= m g \,v_{5}(t)-im \Omega ^2 \hbar \,v_{7}(t)+i(\hbar/m)\,v_{6}(t), \\
%        \dot{v}_2(t) &=  -\theta  \hbar  \left(1+\epsilon  \sin ( \omega_Gt)\right)\,v_{5}(t)+  mg\,v_{10}(t), \\
%        \dot{v}_3(t) &= -\theta  \hbar   \left(1+\epsilon  \sin (\omega_Gt)\right)\,v_{10}(t), \\
%        \dot{v}_4(t) &=  mg\,v_{8}(t)+ m \Omega ^2 \,v_{5}(t), \\
%        \dot{v}_5(t) &=2  mg \,v_{7}(t) -(1/m)\,v_{4}(t), \\
%        \dot{v}_6(t) &= m \Omega ^2 \,v_{8}(t), \\
%        \dot{v}_7(t) &= -(1/m)\,v_{8}(t), \\
%        \dot{v}_8(t) &= 2 m \Omega ^2 \,v_{7}(t)-(2/m)\,v_{6}(t), \\
%        \dot{v}_9(t) &= -  \left(1+\epsilon  \sin ( \omega_Gt)\right) \\
%        &\phantom{=}- \theta  \hbar \left(1+\epsilon  \sin (\omega_Gt)\right)\,v_{8}(t) +m \Omega ^2 \,v_{10}(t), \\
%        \dot{v}_{10}(t) &= -(1/m)\,v_{9}(t)-2 \theta  \hbar \left(1+\epsilon\sin (\omega_Gt)\right)\,v_{7}(t).
%    \end{aligned}
%\end{equation}
%We thus have ten linear differential equations for the components of the quantum Fisher information vector. This is in contrast to the ten nonlinear equations if we were using the Wei-Norman expansion. 

Solving these equations, with initial condition $v_j(0)=0$, gives solution
\begin{equation}
    \begin{aligned}
        v_2(t) &= \frac{gG_0}{\Omega^2\omega_G}\left(1+\frac{\omega_G^2\cos(\Omega t)-\Omega^2\cos(\omega_Gt)}{\Omega^2-\omega_G^2}\right), \\
        v_3(t) &= \frac{\hbar G_0^2}{4m\Omega\omega_G(\Omega^2-\omega_G^2)^2}\left\{(\Omega^2-\omega_G^2)  \left[4\left(\omega_G\sin(\Omega t) \right.\right.\right. \\
        &\phantom{=\hspace{2em}}\left.\left.\left.-\Omega\sin(\omega_Gt)\right)-\theta\Omega\left(2\omega_Gt+\sin(2\omega_Gt)\right)\right]\right. \\
        &\phantom{=\hspace{1em}}\left.+4\theta\omega_G\Omega\left[\Omega\cos(\omega_Gt)\sin(\Omega t)\right.\right. \\
        &\phantom{=\hspace{2em}}\left.\left.-\omega_G\cos(\Omega t)\sin(\omega_Gt)\vphantom{\Omega^2}\right]\right\}, \\
        v_9(t) &= G_0\frac{\omega_G\sin(\omega_Gt)-\Omega\sin(\Omega t)}{\Omega^2-\omega_G^2}, \\
        v_{10}(t) &= \frac{G_0}{m}\frac{\cos(\omega_Gt)-\cos(\Omega t)}{\Omega^2-\omega_G^2}.
    \end{aligned}
\end{equation}

\subsubsection{Time-dependent mechanical frequency}
Let us now consider a harmonic oscilation to the mechanical frequency:
\begin{equation}
    \Omega(t)^2=m\left(\Omega_0^2+\theta\cos(\Omega_mt)\right).
\end{equation}
As with the dynamics, this case will prove to be more troublesome than the other two. The parameter $\theta$ is encoded in the $x^2$ term of the Hamiltonian. Unfortunately the ideal generated by this term coincides with the whole Lie algebra, so we cannot reduce the dimension in this case. Instead we must consider evolution of the whole algebra:
\begin{align}
    \dot{v}_{1}(t)&= mg\,v_{5}(t)+i \hbar  v_{6}(t) \\
    &\phantom{=}-i\hbar m^2 v_{7}(t) \left(\theta  \cos ( \Omega_mt)+\Omega_0^2\right)/m,\nonumber \\
    \dot{v}_{2}(t)&= mg\,v_{10}(t)-\hbar G\, v_{5}(t), \\
    \dot{v}_{3}(t)&=- \hbar  G\,v_{10}(t), \\
    \dot{v}_{4}(t)&=m \left(g\,v_{8}(t)+v_{5}(t) \left(\theta  \cos ( \Omega_m t)+\Omega_0^2\right)\right), \\
    \dot{v}_{5}(t)&=2  mg\,v_{7}(t)-v_{4}(t)/m, \\
    \dot{v}_{6}(t)&=m\,v_{8}(t) \left(\theta  \cos ( \Omega_m t)+\Omega_0^2\right) \\
    &\phantom{=}+m \cos ( \Omega_mt)/(2 \hbar),\nonumber \\
    \dot{v}_{7}(t)&=-v_{8}(t)/m, \\
    \dot{v}_{8}(t)&=2 m\,v_{7}(t) \left(\theta  \cos (\Omega_mt )+\Omega_0^2\right)-2 v_{6}(t)/m, \label{eq:VaryingOmegaFIV8}\\
    \dot{v}_{9}(t)&=m\,v_{10}(t) \left(\theta  \cos (\Omega_mt )+\Omega_0^2\right)- \hbar G\,v_{8}(t), \\
    \dot{v}_{10}(t)&=-\left(2 m\hbar G \,v_{7}(t)+v_{9}(t)\right)/m. 
\end{align}
 Unfortunately this system cannot be solved analytically, thus as in \cite{schneiter_optimal_2020} we will have to follow a perturbative approach. 

We can see that $v_6,v_7,v_8$ form a closed set of equations, which all other equations depend on. Thus we must first solve these, and then substitute them into the other equations. There are various possible approaches, depending on which regime we want to operate in. For example if the perturbation is small, we can note that these have an exact solution when $\theta=0$:
\begin{equation}
    \begin{aligned}
        v_6(t) &= m\frac{\Omega_0\Omega_m\sin(2\Omega_0t)+\left(2\Omega_0^2-\Omega_m^2\right)\sin(\Omega_mt)}{2\hbar\Omega_m\left(4\Omega_0^2-\Omega_m^2\right)}, \\
        v_7(t) &= \frac{2\Omega_0\sin(\Omega_mt)-\Omega_m\sin(2\Omega_0t)}{2\hbar \Omega_m\Omega_0m\left(4\Omega_0^2-\Omega_m^2\right)}, \\
        v_8(t) &= \frac{1}{\hbar}\frac{\cos(2\Omega_0t)-\cos(\Omega_mt)}{4\Omega_0^2-\Omega_m^2}.
    \end{aligned}
\end{equation}
We can then expand the dynamics to first order in $\theta$. We can also expand around some resonance condition between $\Omega_m$ and $\Omega_0$, as was done in \cite{schneiter_optimal_2020}.

\end{document}